\documentclass[
3p,
sort&compress,
]{elsarticle}
\usepackage[bookmarks=false]{hyperref}
\usepackage{setspace}
\usepackage[normalem]{ulem}
\usepackage{multirow, makecell, booktabs}%
\usepackage{algorithm}
\usepackage{xcolor}
\usepackage[noEnd=false, italicComments=false, commentColor=gray!100, rightComments=true]{algpseudocodex}
\usepackage{physics2, fixdif, derivative}
\usepackage{mathtools}
\usepackage{mdframed}
\usepackage{xspace}
\usepackage{lineno}

\usephysicsmodule{braket}
\algnewcommand{\Ifoneline}[2]{\State \algorithmicif\ #1: #2}
\makeatletter
\renewcommand{\ALG@beginalgorithmic}{\footnotesize} 
\makeatother

\usepackage{siunitx}
\DeclareSIUnit{\formulaunit}{\text{f.u.}}
\DeclareSIUnit{\bohrmagneton}{\mu\textsubscript{B}}
\DeclareSIUnit{\angstromcubic}{\text{\AA}\textsuperscript{3}}

\renewcommand{\vec}[1]{\mathbf{#1}}

\def\Psiqm{{\Psi_{\vec{q}m}}}
\def\vxc{V^{\rm xc}}
\def\vxcqsgw{V^\mathrm{xc}_{\text{\scriptsize QSGW}}}
\def\vxclda{V^\mathrm{xc}_{\text{\scriptsize LDA}}}
\def\Psikn{{\Psi_{\vec{k}n}}}
\def\Psiqn{{\Psi_{\vec{q}n}}}
\def\iDelta{{\it \Delta}}
\def\H0{H^0}

\newcommand{\req}[1]{\mbox{Eq.~(\ref{#1})}}
\newcommand{\ecalj}{\texttt{ecalj}\xspace}

\newcommand{\nIBZ}{n_\mathrm{IBZ}}
\newcommand{\nBZ}{n_\mathrm{BZ}}
\newcommand{\nocc}{n_\mathrm{occ}}
\newcommand{\nunocc}{n_\mathrm{unocc}}
\newcommand{\ncore}{n_\mathrm{core}}
\newcommand{\nall}{n_\mathrm{all}}
\newcommand{\npb}{n_\mathrm{pb}}
\newcommand{\ncpb}{n_\mathrm{Cpb}}
\newcommand{\nmb}{n_\mathrm{mb}}
\newcommand{\nqp}{n_\mathrm{qp}}
\newcommand{\nspin}{n_\mathrm{spin}}
\newcommand{\niomega}{n_\omega^\mathrm{I}}
\newcommand{\nomega}{n_\omega}
\newcommand{\Order}{\mathcal{O}}

\algnewcommand{\LeftState}[1]{\Statex \hspace{\algorithmicindent}#1}
\begin{document}

\begin{frontmatter}
\title{Efficient implementation of the quasiparticle self-consistent $GW$ method on GPU}
\author[1]{Masao Obata\corref{cor1}}
\ead{obata@cphys.s.kanazawa-u.ac.jp}

\author[2,3]{Takao Kotani}
\author[1,3]{Tatsuki Oda}

\cortext[cor1]{Corresponding author}
\affiliation[1]{
  organization = {Graduate School of Natural Science and Technology, Kanazawa University},
  addressline = {Ishikawa},
  postcode = {920-1192},
  city = {Kanazawa}, 
  country = {Japan}
}

\affiliation[2]{
  organization = {Department of Engineering, Tottori University},
  addressline = {Tottori},
  postcode = {680-8552},
  city = {Tottori}, 
  country = {Japan}
}

\affiliation[3]{
  organization = {Center for Spintronics Research Network, Osaka University},
  addressline = {Toyonaka},
  postcode = {560-8531},
  city = {Osaka},
  country = {Japan},
}

\begin{abstract}
We have developed a multi-GPU version of the quasiparticle self-consistent $GW$ (QSGW), a cutting-edge method for describing electronic excitations in a first-principles approach.
While the QSGW calculation algorithm is inherently well-suited for GPU computation due to its reliance on large-scale tensor operations, achieving a maintainable and extensible implementation is not straightforward.
Addressing this, we have developed a GPU version within the \texttt{ecalj} package, utilizing module-based programming style in modern Fortran. This design facilitates future development and code sustainability.
Following the summary of the QSGW formalism, we present our GPU implementation approach and the results of benchmark calculations for two types of systems to demonstrate the capability of our GPU-supported QSGW calculations.
\end{abstract}

\begin{graphicalabstract}
  \includegraphics[width=17cm]{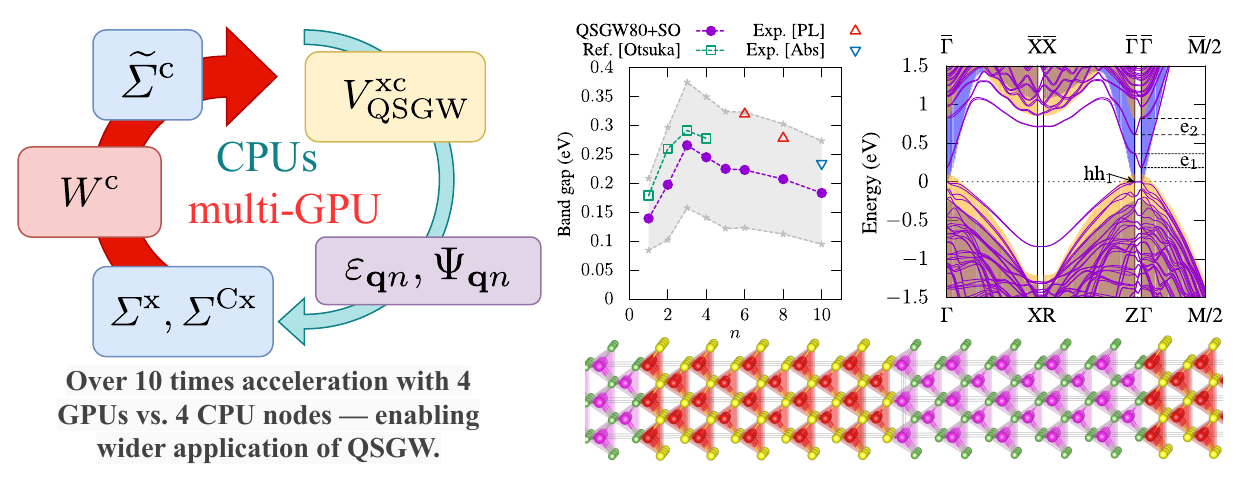}
\end{graphicalabstract}

\begin{highlights}
\item Efficient QSGW calculations leveraging multi-GPU acceleration. \\
  Achieved significant speedup in QSGW calculations utilizing cutting-edge GPU technology; A GPU node achieved 10-fold speedup over four CPU nodes.
  Our implementation ensures maintainability and facilitates future development.
\item Enhanced applicability to large-scale systems. \\
  Introduced a mixed precision approach, achieving a speed increase of more than 3 times while preserving numerical accuracy, thereby enabling further applications to large-scale systems.
\item Accurate electronic structures and new physical insights. \\
The electronic structures obtained from QSGW calculations agree well with the experimental band gaps of InAs/GaSb superlattices, and suggest a possible structural stabilization mechanism in Ni$_2$MnGa, attributed to a valley structure in the density of states near the Fermi level.
  
\end{highlights}

\begin{keyword}
QSGW \sep multi-GPU \sep \ecalj \sep InAs/GaSb \sep Ni$_2$MnGa
\end{keyword}

\end{frontmatter}

\section{Introduction}
\label{sec:introduction}

The first-principles $GW$ approximation (GWA) \cite{aryasetiawan_thegwmethod_1998,Deguchi2016} 
based on the many-body perturbation theory (MBPT) \cite{hedin_effects_1970}
is a powerful method to describe electronic excitations of materials. 
GWA contains not only the exchange effect in the Hartree-Fock theory but also the correlational effect, where we use the dynamical screened Coulomb interaction $W$ instead of the bare Coulomb interaction $v$. 
$W$ is given in the random phase approximation (RPA),
which Hedin and Lundqvist called the linearized time-dependent Hartree approximation \cite{hedin_effects_1970}. 
The quasiparticle (QP) energies in MBPT imply the energies to add or remove an electron from the system, directly corresponding to the experimental spectra obtained from photoemission spectroscopy.
GWA has been employed to calculate QP energies as perturbative corrections to the eigenvalues obtained from the local density approximation (LDA) or the generalized gradient approximation (GGA) within density functional theory (DFT).
Hybertsen and Louie pioneered the first-principles GWA in 1986, demonstrating that it accurately predicts the band gaps of simple semiconductors \cite{Hybertsen1986}. 
This approach is commonly referred to as the one-shot $GW$ method.
However, the one-shot $GW$ method can be problematic in terms of self-consistency.
In particular, it tends to underestimate band gaps significantly in certain materials, such as ZnO and InN \cite{usuda_all-electron_2002,usuda_band_2004, Deguchi2016}.
In cases where there is no band gap at the LDA level, such as Ge, the one-shot $GW$ cannot modify connectivity of eigenvalue dispersion in the Brillouin zone because no off-diagonal matrix elements are included \cite{Schilfgaarde2006}.

To overcome the problem of self-consistency in the one-shot $GW$,
we developed the quasiparticle self-consistent $GW$ (QSGW) method, in which a self-consistent one-body Hamiltonian $H^0$ is obtained within GWA, offering a good independent-particle picture for describing electronic excitations in materials \cite{Schilfgaarde2006, Kotani2007}. 
QSGW provides an improved description of the band gaps and the effective masses compared to the experimental values \cite{Schilfgaarde2006, Otsuka2017, Deguchi2016}. It also accurately describes the positions and bandwidths of localized 3$d$ orbitals \cite{faleev_all-electron_2004, Fabien2006, Jang2015}. This is in contrast to the cases of LDA/GGA, which often give too extended 3$d$ orbitals. Note that such localization changes cannot be described in the one-shot $GW$, which does not treat changes of eigenfunctions from LDA/GGA. 
The effect of $U$ in the LDA+$U$ method is rather naturally incorporated in QSGW through $W$, which is determined self-consistently within the QSGW cycle.

We had implemented QSGW in the \ecalj package \cite{ecalj, Deguchi2016}.
We can handle even metals because we use the coarser $\vec{k}$ mesh for the self-energy calculation to reduce too much computational effort, rather than the mesh used for the self-consistency of the charge.
Furthermore, we can handle magnetic materials \cite{Yoon2019, Obata2023, Jakub2024}.
Since a self-energy interpolation in the Brillouin zone is built-in, we can generate eigenfunctions at any $\vec{k}$ points without requiring additional procedures, such as Wannier interpolations \cite{marzari_maximally_2012}. 
Based on the results of QSGW, linear response functions such as dielectric functions, magnetic susceptibilities, impact ionization energies, etc, can be calculated, enabling a deeper understanding of material properties and facilitating the design of advanced materials for various applications, including electronics, optoelectronics, and spintronics.
This procedure may resolve the intrinsic problems related to the electronic excitations, which are hardly described in LDA. 
For example, we can perform very high-resolution calculations on spin waves \cite{Okumura2019, Okumura2021},
virtually equivalent to the $100 \times 100 \times 100$ division of the Brillouin zone.

Therefore, QSGW is promising for the application of a wide range of materials, not only bulk materials but also interfaces, surfaces, and so on. For such applications, we have two problems. The first is to achieve automatic and robust computation just from atomic positions without manual setting of various computational settings. The second is how to increase the computational speed of QSGW, which requires 100$\sim$1000 times computational effort rather than LDA/GGA.
This article focuses on the latter problem, while the first point is described elsewhere \cite{Kotani2014,takano2025}. 

In this paper, we have implemented the multiple graphics processing unit (GPU) version of QSGW in \ecalj to increase the computational speed. 
Although recent studies have explored QSGW calculations accelerated by GPUs, detailed descriptions of implementation strategies and performance evaluations are still lacking \cite{Ness2024}.
Furthermore, our implementation offers advantages due to the choice of basis set used to represent the eigenfunctions. Specifically, we have implemented QSGW on the PMT method \cite{Kotani2014}, which employs a mixed basis set consisting of both Muffin-tin orbitals (MTOs) and augmented plane waves (APWs) \cite{Kotani2010}.
The use of an appropriate basis set reduced the need for manual tuning of basis-set parameters \cite{Kotani2014}.
The PMT method enables us to treat systems involving vacuum regions, such as slab supercell models \cite{Sakakibara2020}, which are difficult to handle using MTOs alone.
Our implementation is written in an object-oriented style using Fortran modules, making the code structure transparent and maintainable. The source code is publicly available on GitHub \cite{ecalj}.

After showing the computational algorithm in QSGW with key formulas in Sec.~\ref{sec:theory}, 
we will explain how to implement QSGW on multiple GPUs in Sec.~\ref{sec:multigpu}.
In Sec.~\ref{sec:benchmark}, we demonstrate the performance of the implementation for two examples,
Type-II superlattice InAs/GaSb and the ferromagnetic shape-memory alloy Ni$_2$MnGa. 
Both are essential materials from the perspective of practical applications.

\section{Theory and Algorithm to perform QSGW} \label{sec:theory}
\subsection{Formulation of QSGW}
QSGW is a method to determine the one-body Hamiltonian $H^0$ of the independent-particle picture of electronic excitation \cite{Kotani2007}. $H^0$ gives eigenvalues and eigenfunctions of QPs.
Theoretically, the existence of $H^0$ of QPs is supported by Landau's Fermi liquid theory \cite{Landau1957}, where QPs are considered to be the fundamental excitations of electronic systems. Note that QPs can be well illustrated for insulators with band gaps: we have well-defined quasiparticles with infinite lifetime below the impact ionization threshold \cite{Kotani2010, fujita_analysis_2017}, as long as we neglect phonons. This is because there are no decay channels for such QPs \cite{Sakakibara2020}. 
QPs represent the dressed electrons propagating in materials.
Strictly speaking, QPs are defined well only below the threshold. In this sense, $H^0$ is just a theoretical construction to calculate physical quantities on top of $H^0$.

Let us start from the basics of GWA. 
Although any $H^0$ can be chosen, we start from the Kohn-Sham Hamiltonian that incorporates the exchange-correlation potential $\vxc$ within LDA.
From such $H^0$, we have a set of eigenvalues
$\varepsilon_{\vec{k}n}$ and eigenfunctions $\Psi_{\vec{k}n}(\vec{r})$ that satisfy
$H^0 \Psi_{\vec{k}n}(\vec{r})= \varepsilon_{\vec{k}n} \Psi_{\vec{k}n}(\vec{r})$.
GWA gives the self-energy as $\varSigma(\vec{r},\vec{r}',t-t') =  iG(\vec{r},\vec{r}',t-t') W(\vec{r},\vec{r}',t-t'+\delta)$,
where the infinitesimal positive $\delta$ is required to ensure that the exchange term is correct. 
In $\omega$ space, we have $\varSigma(\vec{r},\vec{r'},\omega)$ as
\begin{align}
  \varSigma(\vec{r},\vec{r}',\omega) = \frac{i}{2\pi} \int G(\vec{r},\vec{r}',\omega-\omega') W(\vec{r},\vec{r}',\omega') e^{-i \delta \omega'} d \omega',
\end{align}
where the Green's function $G$ and the dynamical screened Coulomb interaction $W$ are made of $\varepsilon_{\vec{k}n}$ and $\Psi_{\vec{k}n}(\vec{r})$.
This $\varSigma(\vec{r},\vec{r'},\omega)$ plays the role of $\vxc_{\rm LDA}$ in LDA in determining QPs.
In other words, we obtain the one-body Green's function
\begin{align}
  G = \frac{1}{\omega-(H^0 + \varSigma - \vxc)} 
  \label{eq:gwa}
\end{align}
in GWA. We can apply GWA to $H^0$ that contains any $\vxc$. 
As discussed in Sec.~\ref{sec:introduction}, the one-shot $GW$ evaluates only the diagonal part of
$\varSigma - \vxc$ as the correction of the eigenvalues.
The one-shot $GW$ is often referred to as $G_0W_0$.
It is important to note the difference in purpose between one-shot $GW$ and QSGW.
The one-shot $GW$ is to evaluate the difference between quasiparticle energies and eigenvalues in DFT, while QSGW is to determine the one-body Hamiltonian $\H0$.

As discussed in Sec.~\ref{sec:introduction}, a problem with the one-shot $GW$ is its lack of self-consistency.
For example, in systems such as the InAs/GaSb superlattice, which we will use as a sample in the benchmark section \ref{sec:benchmark}, we expect some corrected charge transfer between the layers if the electron density is evaluated using the corrected $G$ in GWA.
To take into account such a corrected charge transfer, a new theoretical framework will be required beyond the one-shot $GW$.
This can be achieved by self-consistency; we should employ the scheme of self-consistent perturbation theory.
Specifically, we need to choose the starting $H^0$ so that the perturbation correction $\varSigma - \vxc$ is minimized. This is the idea of QSGW originally introduced in 2004 \cite{faleev_all-electron_2004}. In QSGW, we will repeat the self-consistency cycle until $\varSigma - \vxc$ is minimized.

\begin{figure}[tb]
  \centering
  \includegraphics[width=8.4cm]{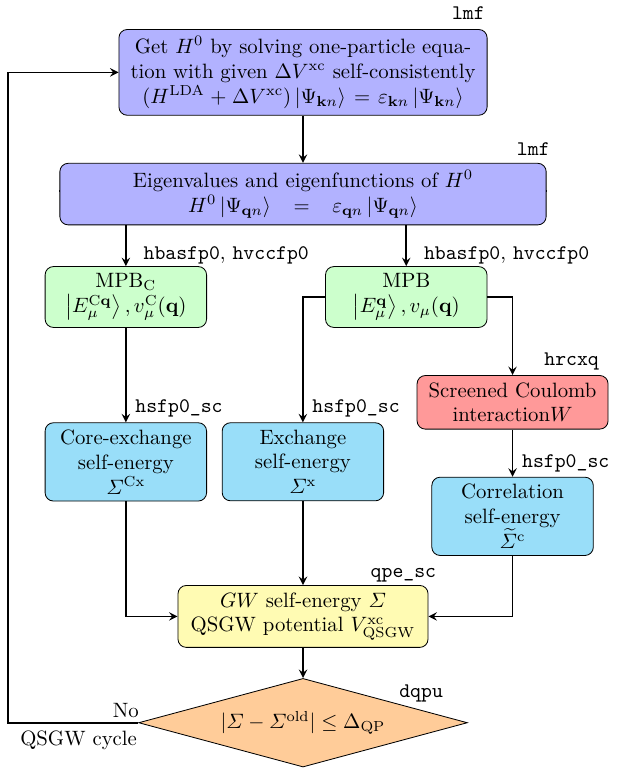}
  \caption{Self-consistent cycle of QSGW. 
    The name of the program that performs each task is noted at the top right corner of the task box.
    The top \texttt{lmf}, which solves the one-particle equation with given $\Delta V^\mathrm{xc}$, performs self-consistent calculations.
    The $\Delta V^\mathrm{xc}$ is set to zero in the initial loop.
    The QSGW cycle continues iteratively, checking if the difference between the new and old diagonal terms of self-energy is below a certain threshold $\Delta_\mathrm{QP}$. If not, the process repeats.
}
  \label{fig:qsgw-cycle}
\end{figure}

To achieve a closed-form self-consistency cycle in QSGW, we need to remove the energy dependence of $\varSigma$ in GWA in Eq.~(\ref{eq:gwa}). That is, we replace $\varSigma$ with the static self-energy $\vxcqsgw$. Thus, in practice, the QSGW iteration cycle is given as in Fig.~\ref{fig:qsgw-cycle}.
At the top of Fig.~\ref{fig:qsgw-cycle}, for a given $\iDelta \vxc$ keeping, we determine 
$\H0$ such that the electron density $n(\vec{r})$ is self-consistently obtained.
$\H0$ contains an extra term, a non-local static potential
$\iDelta \vxc= \vxcqsgw -\vxclda$, which is the difference in the exchange-correlation term between QSGW and LDA.
At the beginning of the iteration cycle, we set $\iDelta \vxc$ to zero.
In the next step, we generate the eigenvalues and eigenfunctions of $H^0$.
Subsequently, after some steps (Green, Red, and Blue boxes in Fig.~\ref{fig:qsgw-cycle}), we finally have $\iDelta V^{\rm xc}$.
The four computationally-expensive steps for calculating $\vxcqsgw$, depicted in red and blue boxes, are as follows.
\begin{mdframed}[userdefinedwidth=8cm, align=center]
  \begin{itemize}
    \setlength{\parskip}{0pt}
    \setlength{\itemsep}{0pt}
    \item[1.] Core-exchange self-energy $\varSigma^\mathrm{Cx}$
    \item[2.] Exchange self-energy $\varSigma^\mathrm{x}$
    \item[3.] The dynamical screened Coulomb interaction $W$ \text{\ (red box in Fig.~\ref{fig:qsgw-cycle})}
    \item[4.] Static version of correlation self-energy $\widetilde{\varSigma}^\mathrm{c}$
  \end{itemize}
\end{mdframed}
We need $W$ to calculate the static version of the correlation self energy $\widetilde{\varSigma}^\mathrm{c}$.
The matrix elements of the correlation self-energy $\widetilde{\varSigma}^\mathrm{c}$ is written as
\begin{align}
  \widetilde{\varSigma}^\mathrm{c}_{nm}(\vec{k}) &
  = \braket[3]{\Psi_{\vec{k} n}}{ 
  {\rm Re}\left[
  {\frac{ {\varSigma}^{\rm c}(\varepsilon_{\vec{k}n}) + {\varSigma}^{\rm c}(\varepsilon_{\vec{k}m}) }{2}}
  \right]}{\Psi_{\vec{k} m}},  \label{eq:vxc}
\end{align}
where Re means taking the Hermitian part. 
To eliminate the $\omega$ dependence of $\varSigma^{\rm c}$, 
we use \req{eq:vxc}, which is one of the reasonable choices \cite{Kotani2007}.
Together with other terms, we obtain
$\vxcqsgw =\varSigma^\mathrm{Cx}+\varSigma^\mathrm{x}+ \widetilde{\varSigma}^\mathrm{c}$ in the expansion of $\Psi_{\vec{k} n}$ of $H^0$. 
After self-consistency is achieved, $\{\Psi_{\vec{k} n}\}$ used for expansion in \req{eq:vxc} is in agreement with $\{\Psi_{\vec{k} n}\}$ generated in the next step of the iteration cycle.

We divide all electrons into core and valence electrons by the default setting in \ecalj.
Except for cores, we treat all electrons as valence electrons. 
We use the frozen-core approximation; core electron densities are fixed to be those calculated for spherical atoms in the top box of Fig.~\ref{fig:qsgw-cycle}. 
However, we generate core eigenfunctions restricted within Muffin-tin (MT) regions for the calculation of $\varSigma^\mathrm{Cx}$.
We neglect contributions of cores to $W$, based on the examination in Ref.~\cite{Kotani2007}.
Thus, the cores affect the valence electrons only through $\varSigma^\mathrm{Cx}$, in addition to their frozen electron density.
This treatment of core electrons is justified by comparing cases in which the cores are treated as frozen cores or as valence electrons in the local orbital method.

An empirical correction can be introduced in QSGW to improve agreement with experimental values.
This is because QSGW gives a systematic overestimation of exchange effects \cite{Deguchi2016}. 
This overestimation is reasonable considering the nature of QSGW because the vertex corrections \cite{Shishkin2007, Chen2015} are not considered.
Since the vertex correction is expected to reduce the energies of the intermediate states due to the correlational motions between electrons and holes, we anticipate that the actual effective interaction between electrons will be smaller than that evaluated in QSGW.
The electrons should be polarized more easily than they would be without the vertex correction.
To account for this effect of vertex correction in a simple manner, we can use QSGW80, a hybrid with a mixing ratio of 80\% QSGW and 20\% LDA, rather than 100\% QSGW (QSGW100).
We have investigated the capability of QSGW80, demonstrating a universal improvement in agreement with experimental data on band gaps and effective masses in semiconductors and insulators \cite{Deguchi2016}.
For QSGW80, we use $0.8 \times \iDelta \vxc$ instead of $\iDelta \vxc$.

Hereafter, we will show detailed equations for the four steps, followed by the representation of eigenfunctions in the MT division of space.

\subsection{Representation of eigenfunctions in the MT division of space}
Our method is based on an all-electron electronic structure calculation in the PMT method \cite{Kotani2002, Kotani2010}, implemented in the \ecalj package \cite{ecalj}.
The PMT is a linearized mixed-basis method that utilizes not only MTOs but also APWs to represent the eigenfunctions, employing the MT division of the space.
In this context, the PMT is a combination of linearized MTO (LMTO) and linearized APW (LAPW) methods.
The APW effectively represents the smooth behavior of eigenfunctions in the interstitial region, whereas the MTO effectively represents the eigenfunctions just outside of MTs.
In \ecalj, the parameters for defining MTOs are determined in a simple manner.
This is because APWs, with a cutoff energy of approximately \SI{3}{Ry}, can effectively complement the expansion of eigenfunctions in the interstitial region, thereby avoiding the need to tune the parameters of MTOs \cite{Kotani2010}.
Since the MTO part can expand the eigenfunctions near the MT boundaries very well, we can roughly take the cutoff energy of the APWs as that of the computed eigenfunctions relative to some averaged level of interstitial potential.
Exactly speaking, we also use local orbitals \cite{singh_introduction_2006}. 
The local orbitals are important in cases; for example, we represent $3d$ orbitals with local orbitals for Ga, $2p$ for Na as well. 

For given $\H0$, we obtain eigenfunctions $\Psi_{{\vec{k}}n}(\vec{r})$ expanded in the PMT bases.
We re-expand $\Psi_{\vec{k}n}(\vec{r})$ in the MT space of division as the sum of the augmentation parts within the MTs and the plane wave (PW) parts in the interstitial region as follows:
\begin{align}
\Psikn(\vec{r})
= \sum_{\vec{R} u}  \alpha^{{\vec{k}}n}_{\vec{R} u} \varphi^{\vec{k}}_{\vec{R} u}(\vec{r})
+ \sum_{\vec{G}}  \beta^{{\vec{k}}n}_{\vec{G}} P^{\vec{k}}_{\vec{G}}({\vec{r}}),
\label{def:psiexp}
\end{align}
where the interstitial plane wave (IPW) is defined as
\begin{align}
  P^{\vec{k}}_{\vec{G}}({\vec{r}}) =
\begin{cases}
0    & \text{if}~\vec{r} \in \text{any MT} \\
  \exp(i ({\vec{k+G}})\cdot \vec{r})& \text{otherwise}
\end{cases}
\label{eq:defpg}
\end{align}
and $\varphi^{\vec{k}}_{\vec{R} u}(\vec{r})$ are Bloch sums of the atomic functions $\varphi_{\vec{R} u}(\vec{r})$ defined within the MT at $\vec{R}$,
\begin{align}
\varphi^{\vec{k}}_{\vec{R} u}({\vec{r}}) \equiv  \sum_{\vec{T}} \varphi_{\vec{R} u}(\vec{r-R-T}) \exp(i \vec{k\cdot T}).
\end{align}
$\vec{T}$ and $\vec{G}$ are lattice translation vectors in real and reciprocal spaces, respectively.

\subsection{Overview of QSGW calculations} \label{sec:overview_of_qsgw}

In our method to perform GWA, we need not only the basis set to expand eigenfunctions but also the basis set to expand the product of eigenfunctions.
The basis set is referred to as the mixed product basis set (MPB) $\{M^{\vec{k}}_I({\vec{r}})\}$, introduced by Kotani in Ref.~\cite{Kotani2002}.
The MPB consists of the product basis (PB) within MTs, by Aryasetiawan and Gunnarsson \cite{Aryasetiawan1994}, and the IPW in the interstitial region.
Since $\{M^{\vec{k}}_I({\vec{r}}) \}$ contains IPWs that are not orthogonal, we have an overlap matrix $O^{\vec{k}}_{IJ}= \braket{M^{\vec{k}}_{I}}{M^{\vec{k}}_J}$.
We introduce an orthonormal MPB $\{ E^{\vec{k}}_\mu \}$ that diagonalizes the Coulomb matrix for convenience.
An orthonormal MPB was originally introduced by Friedrich, Bl\"ugel, and Schindlmayr in Ref.~\cite{Friedrich2010}.
Let the transformation coefficients from $\{ M^{\vec{k}}_I \}$ to $\{ E^{\vec{k}}_\mu\}$ be $z^{\vec{k}}_{\mu I}$, and let $v_\mu({\vec{k}})$ be the eigenvalue of $v(\vec{k})$, i.e., $\ket{E^{\vec{k}}_\mu} = \sum_I z^{\vec{k}}_{\mu I} \ket{M^{\vec{k}}_I}$, $v(\vec{k}) \ket{E^{\vec{k}}_\mu} = v_\mu({\vec{k}}) \ket{E^{\vec{k}}_\mu}$.
The values of $z^{\vec{k}}_{\mu I}$ and $v_\mu(\vec{k})$ are obtained by solving the generalized eigenvalue equation, $\sum_J (v_{IJ}^{\vec{k}} - v_{\mu}(\vec{k}) O^{\vec{k}}_{IJ} ) z^{\vec{k}}_{\mu J} = 0$, where $v_{IJ}^{\vec{k}} = \braket[3]{M^{\vec{k}}_{I}}{v}{M^{\vec{k}}_J}$.
By using these, the Coulomb interaction can be represented by matrix elements as follows: 
\begin{align}
  v(\vec{k})=\sum_{\mu} \ket{E^{\vec{k}}_\mu} {v_\mu(\vec{k})} \bra{E^{\vec{k}}_\mu}. \label{eqvcoue}
\end{align}

Equation~\eqref{eqvcoue}, as introduced in Ref.~\cite{Friedrich2010}, is used in the all-electron full-potential GWA.
This corresponds to the representation in the plane wave expansion of the Coulomb interaction, given by $v(\vec{k}+\vec{G},\vec{k}+\vec{G}')=\frac{4 \pi \delta_{\vec{G} \vec{G}'}}{|\vec{k}+\vec{G}|^2}$.
The case of $\mu=1$ corresponds to the largest eigenvalue of $v_{\mu}$, which is $v_{\mu=1}(\vec{k})$ $\sim \frac{4 \pi}{|\vec{k}|^2}$.
We handle this divergence numerically by replacing $1/r$ with $\exp(- \kappa_r r)/r$, where $\kappa_r$ is small enough
(\SI{1d-5}{bohr^{-1}}, usually).
By introducing $\{ E_\mu^{\vec{k}} \}$ as the new MPB, our $GW$ calculations are formulated using the matrix elements $Z$ in terms of $E_\mu^{\vec{k}}$ 
and the wave functions $\Psi$ which are written as follows:
\begin{align}
 Z^{\vec{q-k}n', \vec{q}n}_{\vec{k}\mu} = \braket{E^{\vec{k}}_{\mu} \Psi_{\vec{q-k}n'}}{\Psi_{\vec{q}n}}. \label{eq:defz}
\end{align}

We expect that the set of mixed product bases $\{M^{\vec{k}}_J\}$ (or $\{E^{\vec{q}}_\mu\}$, equivalently) must have the ability to expand the product of eigenfunctions very well.
However, we need to keep the number of basis functions as small as possible to minimize computational efforts.
Since the size of the $Z^{\vec{q-k} n, \vec{k}n'}_{\vec{q}\mu }$ array can be significantly large, we prepare it on the fly for computations.
The expansion \req{def:psiexp} is designed for the convenience of computing. 
In the following $GW$ calculations, we use $Z^{\vec{k} n ,\vec{q+k n'}}_{\vec{q}\mu}$ with $v_\mu(\vec{k})$, in addition to eigenvalues.

The exchange self-energy is written as
\begin{align}
  \varSigma^\mathrm{x}_{nm}(\vec{q}) &= \braket[3]{\Psiqn}{\varSigma^\mathrm{x}}{\Psiqm}  \nonumber \\
                                     &=-\sum^{\rm BZ}_{{\vec{k}}}  \sum^{\rm  occ}_{n'}  
                                     \sum_{\mu} [Z^{\vec{q-k} n',\vec{q} n}_{\vec{k}\mu }]^* v_{\mu}({\vec{k}}) Z^{\vec{q-k}n' ,\vec{q}m}_{\vec{k}\mu }. \label{eq:sigx}
\end{align}
Replacing occupied valence states in Eq.~(\ref{eq:sigx}) with core states, we can calculate core exchange $\varSigma^\mathrm{Cx}$.
Note that the MPB for $\varSigma^\mathrm{Cx}$ differs from that for valence electrons because the former must represent products of the core and valence eigenfunctions.
Specifically, we need to evaluate $Z^{\vec{q-k}n',\vec{q}n}_{\vec{k}\mu}$, where the intermediate states $\Psi_{\vec{q-k}n'}$ belong to the cores just within MTs.
The screened Coulomb interaction $W_{\mu\nu}(\vec{q}, \omega)$ is calculated from the dielectric function $\epsilon_{\mu\nu}(\vec{q}, \omega)$ and the Lindhard polarization function $\varPi_{\mu\nu}(\vec{q}, \omega)$ i.e,
\begin{align}
  \epsilon_{\mu\nu}(\vec{q}, \omega) &= \delta_{\mu\nu} - \sqrt{v_\mu(\vec{q})} \varPi_{\mu\nu}(\vec{q},\omega) \sqrt{v_\nu(\vec{q})}, \label{eq:dielectricfunction} \\
  W_{\mu\nu}(\vec{q}, \omega) &= \sqrt{v_\mu(\vec{q})} \epsilon_{\mu\nu}^{-1}(\vec{q},\omega) \sqrt{v_\nu(\vec{q})}, \label{eq:screenedcoulomb}
\end{align}
where the $\varPi_{\mu\nu}(\vec{q},\omega)$ is written as
\begin{align}
  \varPi_{\mu \nu}({\vec{q}},\omega)
&= \sum^{\rm BZ}_{\vec{k}} \sum^{\rm occ}_{n} \sum^{\rm unocc}_{n'} 
 \frac{  Z^{\vec{k} n,\vec{q+k}n'}_{\vec{q}\mu } [Z^{\vec{k}n ,\vec{q+k}n'}_{\vec{q}\nu }]^*}{\omega-(\varepsilon_{\vec{q+k} n'}-\varepsilon_{\vec{k} n})+i \delta} 
+ \sum^{\rm BZ}_{\vec{k}} \sum^{\rm  unocc}_{n} \sum^{\rm occ}_{n'}
 \frac{  Z^{\vec{k} n,\vec{q+k} n'}_{\vec{q}\mu} [Z^{\vec{k}n ,\vec{q+k}n'}_{\vec{q}\nu }]^*}{-\omega-(\varepsilon_{\vec{k} n}-\varepsilon_{\vec{q+k} n'})+i \delta}
 \label{eq:polf01} \\
&= \sum^{\rm BZ}_{\vec{k}} \sum^{\rm occ}_{n} \sum^{\rm unocc}_{n'} 
 \frac{  Z^{\vec{k} n,\vec{q+k} n'}_{\vec{q}\mu} [Z^{\vec{k}n ,\vec{q+k} n'}_{\vec{q}\nu}]^*}{\omega-(\varepsilon_{\vec{q+k} n'}-\varepsilon_{\vec{k} n})+i \delta} 
+ \sum^{\rm BZ}_{\vec{k}} \sum^{\rm  occ}_{n} \sum^{\rm unocc}_{n'}
 \frac{  Z^{\vec{-k} n *,\vec{-q-k} n'*}_{\vec{q}\mu} [Z^{\vec{-k} n*,\vec{-q-k}n'*}_{\vec{q}\nu}]^*}{-\omega-(\varepsilon_{\vec{-q-k} {n'}}-\varepsilon_{\vec{-k} n})+i \delta}.
\label{eq:polf02} 
\end{align}
Note that $Z^{\vec{-k} n *,\vec{-q-k} n'*}_{\vec{q}\mu}=
\braket{E^{\vec{q}}_{\mu} \Psi^*_{\vec{-k}n}}{\Psi^*_{\vec{-q-k}n'}}$.
In the second term of Eq.~(\ref{eq:polf02}), we interexchange $n$ and $n'$ of Eq.~(\ref{eq:polf01})
with keeping momentum conservation. 
$\Pi_{\mu\nu}(\vec{q}, \omega)$ can be divided
into time-revesal and anti-time-reversal parts as discussed in Ref.~\cite{Kotani2007}.
If we assume time-reversal symmetry of $H^0$,
$\Psi_{\vec{k} n}^*$ is the degenerated eigenfunciton of $\Psi_{\vec{-k} n}$. 
Then the numerator of Eq.~(\ref{eq:polf02}) at the second term is identical to that of the first term;
thus the antitime-reversal parts disappear. Then we have 
\begin{align}
&\varPi_{\mu \nu}({\vec{q}},\omega)
=\sum^{\rm BZ}_{{\vec{k}}}  \sum^{\rm  occ}_{n} \sum^{\rm  unocc}_{n'}
  Z^{\vec{k} n,\vec{q+k}n'}_{\vec{q} \mu } [Z^{\vec{k}n ,\vec{q+k} n'}_{\vec{q}\nu}]^* \nonumber \\
& \times
\left(\frac{1}{\omega-\varepsilon_{\vec{q}+\vec{k}n'}+\varepsilon_{\vec{k}n}+i \delta}
-\frac{1}{\omega+\varepsilon_{\vec{q+k}n'}-\varepsilon_{\vec{k}n}-i \delta}\right).
\label{eq:polfx}
\end{align}
Hereafter, we assume the time-reversal symmetry.
This definition is the same as that in the Fetter-Walecka textbook \cite{Fetter1971}.
At $\omega=0$, this is a negative definite matrix.
The matrix part $Z^{\vec{k} n,\vec{q+k}n'}_{\vec{q} \mu } [Z^{\vec{k}n ,\vec{q+k} n'}_{\vec{q}\nu}]^*$ in Eq.~(\ref{eq:polfx}) is Hermitian, and $\varPi_{\mu \nu}(\vec{q},\omega)=\varPi_{\mu \nu}(\vec{q},-\omega)$ is satisfied for general complex $\omega$
where we can neglect $\pm i \delta$.
In practice, we use the Hilbert transformation to calculate the Hermitian part of $\varPi$ after computing the anti-Hermitian part.
The anti-Hermitian part of $\varPi$, written as ${\rm Im} \varPi$, is written as 
\begin{align}
    {\rm Im} [\varPi_{\mu \nu}(\vec{q},\omega_j)]
                                       &=\frac{\pi}{i}\sum^{\rm BZ}_{\vec{k}}  
                                       \sum^{\rm  occ}_{n} \sum^{\rm  unocc}_{n'}
  Z^{\vec{k} n,\vec{q+k}n'}_{\vec{q}\mu } [Z^{\vec{k}n,\vec{q+k} n'}_\vec{q\nu}]^* 
  w_{n'n}^{\vec{q},\vec{k}}(\omega_j), \label{eq:impolf}
\end{align}
where $w_{n'n}^{\vec{q},\vec{k}}(\omega_j)$ is the tetrahedron weight to accumulate $\mathrm{Im} \varPi$.
This weight is calculated as follows:
\begin{align}
  w_{n'n}^{\vec{q},\vec{k}}(\omega_j)&= \sum_{\vec{k} \in {\rm BZtet}(i_{\rm tet})} \frac{1}{4} w(n',n,\vec{q},\vec{k},i_{\rm tet}, \omega_j) \label{eq:tetw}, \\
    w(n',n,\vec{q},\vec{k},i_{\rm tet},\omega_j)&= \int_{{\rm BZtet}(i_{\rm tet})} d^3k \int_{\omega_{{\rm L}_j}}^{\omega_{{\rm R}_j}}
    \delta(\omega-\varepsilon_{\vec{q+k}n'}+\varepsilon_{\vec{k}n}) d \omega, \label{eq:tetrahedronweeight}
\end{align}
where $\omega_j$ denotes a histogram bin $[\omega_{{\rm L}_j},\omega_{{\rm R}_j}]$ for $\omega$ along the real axis to accumulate the imaginary part.
We employ a logarithmic mesh for $\omega_j$ such as $\omega_j= 10^{-5} \times 1.03^j$ Ry, resulting in $\sim$100 mesh points to cover all excitation energies.
We evaluate the sum for $\vec{k}$ of the right-hand side of \req{eq:impolf} using a tetrahedron method for the polarization function \cite{Kotani2014, Rath1975}. 
Note that the products $Z^{\vec{k}n,\vec{q+k}n'}_{\vec{q}\mu} [Z^{\vec{k}n,\vec{q+k}n'}_{\vec{q}\nu}]^*$ are the positive definite matrix.
The tetrahedron method is crucial for performing accurate $GW$ calculations efficiently.
We expect the correct sum rule to be satisfied within the space spanned by the PMT bases, which are used to expand eigenfunctions.
The real (Hermitian) part is obtained through the Hilbert transformation as
\begin{align}
  {\rm Re} [\varPi_{\mu \nu}(\vec{q},\omega_i)] &= \sum_j K_{ij} {\rm Re} [\varPi_{\mu \nu}(\vec{q},\omega_j)],
\label{eq:hilbert}
\end{align}
where $K_{ij}$ is the Hilbert transformation matrix, a discretized version of $1/(\omega_i - \omega_j)$, 
generated in advance with taking into account the size of histogram bins.
By construction, our manner to obtain $\varPi_{\mu \nu}(\vec{q},\omega)$ satisfies 
the Kramers-Kronig relation very well.
Then we obtain $\varPi_{\mu \nu}(\vec{q},\omega_i)$ as the sum of their imaginary part and real part.
When we have no time-reversal symmetry, we accumulate not only the positive energy part of 
Eq.~(\ref{eq:impolf}), but also the negative energy part for occupied $n'$ and unoccupied $n$ in Eq.~(\ref{eq:polf01}). 
As an approximation, we may use Eq.~(\ref{eq:polfx}) even for systems without time-reversal, 
however, we may need some careful examination.

The correlation part of the screened Coulomb interaction 
$W_{\mu\nu}^{\rm c}(\vec{k}, \omega)=W_{\mu\nu}(\vec{k},\omega)-v_\mu(\vec{k})\delta_{\mu\nu}$
is given by the matrix inversion of Eq.~(\ref{eq:dielectricfunction}) as
\begin{align}
  W_{\mu\nu}^{\rm c}(\vec{k}, \omega)&=\sqrt{v_\mu(\vec{k})}(\varepsilon^{-1}_{\mu\nu}(\vec{k},\omega)
  -\delta_{\mu\nu})\sqrt{v_\nu(\vec{k})}. \label{eq:wc}
\end{align}
With $W_{\mu\nu}^{\rm c}(\vec{k}, \omega)$, we have a formula for the correlation self energy as
\begin{align}
\varSigma^\mathrm{c}_{n,m}(\vec{q},\omega) = 
\sum_{\vec{k},n'\mu,\nu} \int_{-\infty}^\infty  \frac{i d \omega'}{2 \pi}
\frac{
[Z^{\vec{q-k}n',\vec{q}n}_{\vec{k}\mu  }]^*
W^{\rm c}_{\mu\nu}(\vec{k},\omega')
Z^{\vec{q-k}n',\vec{q}m}_{\vec{k}\nu  } 
e^{-i \delta \omega'}
}
{\omega-\omega'-\varepsilon_{\vec{q}-\vec{k} n'}\pm i \delta}.
\label{sigmann}
\end{align}
Here, we use $-i \delta$ for the occupied states of ${\vec{q}\!-\!\vec{k} n'}$ and $+i \delta$ for the unoccupied states.
The factor $e^{-i\delta\omega'}$ is neglected due to the behavior of $W^\mathrm{c}(\vec{k},\omega')$ at $\omega' \rightarrow \pm\infty$.

In QSGW, we have to calculate $\varSigma^{\rm c}_{nm}(\vec{q},\varepsilon_{\vec{q} n})=\langle \Psiqn |\varSigma^{\rm c}(\vec{q}, \varepsilon_{\vec{q} n})|\Psiqm \rangle$ as discussed in Eq.~(\ref{eq:vxc}).
The integration contour for $\omega'$ in Eq.~(\ref{sigmann}) is deformed to the one shown in Fig.~15 of \cite{Kotani2007}.
We evaluate $\varSigma^{\rm c}_{nm}(\vec{q},\varepsilon_{\vec{q} n})$ as the sum of the imaginary-axis-contour contribution $I^{\rm c}(\vec{q}, \varepsilon_{\vec{q}n})$ and the real-axis-contour (or pole) contribution $R^{\rm c}(\vec{q}, \varepsilon_{\vec{q}n})$, that is,
\begin{align}
\varSigma^{\rm c}_{nm}(\vec{q},\varepsilon_{\vec{q} n}) = I_{nm}^{\rm c}(\vec{q}, \varepsilon_{\vec{q}n}) + R_{nm}^{\rm c}(\vec{q}, \varepsilon_{\vec{q}n}).
\end{align}
Here $I^{\rm c}(\vec{q}, \varepsilon_{\vec{q}n})$ is written as
\begin{align}
  I_{nm}^{\rm c}(\vec{q}, \varepsilon_{\vec{q}n}) =& -\sum_{\vec{k},n'}\sum_{\mu,\nu} 
  \int_{-\infty}^{\infty} \frac{d\omega}{2\pi}
\frac{
[Z^{\vec{q-k} n',\vec{q} n}_{\vec{k}\mu}]^*
W^{\rm c}_{\mu\nu}(\vec{k}, i\omega)
Z^{\vec{q-k}n',\vec{q} m}_{\vec{k}\nu } 
} {\varepsilon_{\vec{q} n}- i\omega -\varepsilon_{\vec{q}-\vec{k} n'}}   \nonumber \\
\overset{\text{discretized}}{\simeq}
& -\sum_{\vec{k},n'} \sum_{i}  \sum_{\mu,\nu} 
w_i^\mathrm{I}(\varepsilon_{\vec{q} n}-\varepsilon_{\vec{q}-\vec{k} n'})[Z^{\vec{q-k} n',\vec{q} n}_{\vec{k}\mu}]^*
W^{\rm c}_{\mu\nu}(\vec{k}, i\omega^{\rm I}_i)
Z^{\vec{q-k} n',\vec{q} m}_{\vec{k}\nu}. \label{eq:sigci}
\end{align}
The mesh points $\{i\omega^{\rm I}_i\}$ and the corresponding discretized integration weight 
 $\{w^\mathrm{I}({\it \Delta} E,i\omega^{\rm I}_i)|\text{for } i\omega^{\rm I}_i\}$ satisfy
\begin{align}
\int_{-\infty}^{\infty} \frac{d\omega}{2\pi}
\frac{X(i \omega)}{{\it \Delta} E- i\omega } 
\overset{\text{discretized}}{\simeq}
\sum_{i\omega_i^\mathrm{I}} w^\mathrm{I}({\it \Delta} E,i\omega_i^\mathrm{I}) X(i\omega_i^\mathrm{I}),
\end{align}
where $X(i \omega)$ behaves monotonically as
$W^{\rm c}_{\mu\nu}(\vec{k}, \omega)$ along the imaginary axis: $W^{\rm c}_{\mu\nu}(\vec{k}, \omega)$ is a positive-definite hermitian with monotonic damping for $\omega \to \pm i\infty$. 
To determine $w^\mathrm{I}({\it \Delta} E,i\omega_i^\mathrm{I})$, we use the method given by Aryasetiawan and Gunnarsson \cite{Aryasetiawan1994}.
Since $W^{\rm c}(\omega)=W^{\rm c}(-\omega)$, 
we can reduce the sum for $i \omega^{\rm I}_i $ only for $\omega^{\rm I}_i \ge 0$. 
We typically use only 10 points for $\{\omega^{\rm I}_i\}$.
The contribution $R^{\rm c}(\vec{q}, \varepsilon_{\vec{q}n})$ is written as
\begin{align}
  R_{nm}^{\rm c}(\vec{q},\varepsilon_{\vec{q} n})=& \sum_{\vec{k},n'} \sum_{\mu,\nu}
[Z^{\vec{q-k}n',\vec{q} n}_{\vec{k}\mu}]^*
W^{\rm c}_{\mu\nu}(\vec{k},\varepsilon_{{\vec{q}}n}-\varepsilon_{\vec{q}-\vec{k} n'}) 
Z^{\vec{q-k}n',\vec{q} m}_{\vec{k}\nu } \nonumber \\ 
  \times&[\ \ \ 
    \theta(\varepsilon_{{\vec{q}}n}-E_{\rm F})
    \theta(\varepsilon_{\vec{q} n}-\varepsilon_{\vec{q}-\vec{k} n'}) 
    \theta(\varepsilon_{\vec{q}-\vec{k} n'}-E_{\rm F}) 
    \nonumber \\
        &- 
    \theta(E_{\rm F}-\varepsilon_{\vec{q} n})
    \theta(\varepsilon_{\vec{q}-\vec{k} n'}-\varepsilon_{\vec{q} n})
    \theta(E_{\rm F}-\varepsilon_{\vec{q}-\vec{k} n'})]  \nonumber \\
\overset{\text{discretized}}{\simeq}
 & 
\sum_{\vec{k},n'} \sum_{\omega^{\rm R}_i}  \sum_{\mu,\nu} 
w^\mathrm{R}_i(\varepsilon_{\vec{q} n}-E_{\rm F},\varepsilon_{\vec{q}-\vec{k} n'}-E_{\rm F})
[Z^{\vec{q-k}n',\vec{q}  n}_{\vec{k}\mu}]^*
W^{\rm c}_{\mu\nu}(\vec{k}, \omega^{\rm R}_i)
Z^{\vec{q-k}n',\vec{q} m}_{\vec{k}\nu }. \label{eq:sigcr}
\end{align}
Here, the mesh points $\omega_i^\mathrm{R}$ are the boundaries of the histogram bins $\omega_j$ used in
Eq.~(\ref{eq:tetrahedronweeight}). The corresponding discretized integration weight 
$w_i^\mathrm{R}(\varepsilon_{\vec{q} n}-E_{\rm F},\varepsilon_{\vec{q}-\vec{k} n'}-E_{\rm F})$
is the interpolation coefficients to obtain 
$W^{\rm c}_{\mu\nu}$ at $\varepsilon_{\vec{q} n}-\varepsilon_{\vec{q}-\vec{k} n'}$ 
from $W^\mathrm{c}(\omega_i^\mathrm{R})$. 
In addition, the factor from the step function should be included.
Since we employ the three-point interpolation,
$w_i^\mathrm{R}(\varepsilon_{\vec{q} n}-E_{\rm F},\varepsilon_{\vec{q}-\vec{k} n'}-E_{\rm F})$
are non-zero only at three of the mesh points $\omega_i^\mathrm{R}$.
Note that the lifetime of QP (= the impact ionization rate) due to electron-electron scattering is only included in $R^{\rm c}(\vec{q}, \varepsilon_{\vec{q}n})$ \cite{Kotani2010}.
In the calculation of the static version of the correlation self-energy, $W^\mathrm{c}_{\mu\nu}$ in $R^\mathrm{c}_{nm}$ is hermitized.

Regarding the division $\varSigma^{\rm c}_{nm}(\vec{q},\omega) = R_{nm}^{\rm c}(\vec{q},\omega)+I_{nm}^{\rm c}(\vec{q},\omega)$, only $R_{nm}^{\rm c}(\vec{q},\varepsilon_{\vec{q} n})$ contributes to the lifetime of QPs. 
For the electron at $\vec{q}n$( $\varepsilon_{{\vec{q}}n}>E_{\rm F}$), there is a factor $\theta(\varepsilon_{\vec{q} n}-\varepsilon_{\vec{q}-\vec{k} n'})  \theta(\varepsilon_{\vec{q}-\vec{k} n'}-E_{\rm F})$. When $W^{\rm c}_{\mu\nu}(\vec{k},\varepsilon_{{\vec{q}}n}-\varepsilon_{\vec{q}-\vec{k} n'}) $ has the non-zero imaginary part, $R_{nm}^{\rm c}(\vec{q},\varepsilon_{\vec{q} n})$ includes the decay rate to the lower energy electron $\varepsilon_{\vec{q}-\vec{k} n'}$ with an electron-hole pair. 
The details of discussions about the lifetime of QPs and the impact ionization rate in semiconductors were presented in \cite{Kotani2010}.
Since we use Gaussian broadening at energies $\varepsilon_{\vec{q}-\vec{k} n'}$  for the discretized sum in \req{eq:sigcr}, we should truncate the Gaussian weights by the step functions in \req{eq:sigcr}.

We have two key points not yet explained.
One is the improved offset-$\Gamma$ method \cite{Kotani2007}, which treats the divergence of $W^{\rm c}(\vec{k} \to 0,\omega)$ in Eqs.~(\ref{eq:sigx}) and (\ref{sigmann}).
Instead of $W^{\rm c}(\vec{k} \to 0,\omega)$, we use a non-divergent effective interaction $\overline{W^{\rm c}}(\vec{k}=0,\omega)$ in \req{sigmann}.
This $\overline{W^{\rm c}}(\vec{k}=0,\omega)$ is an average of ${W^{\rm c}}(\vec{k},\omega)$ in $\vec{k}$ within the $\Gamma$ cell, which is defined as the microcell of the BZ containing $\vec{k}=0$ \cite{Kotani2014} at the center.
To obtain $\overline{W^{\rm c}}(\vec{k}=0,\omega)$, we have to calculate ${W^{\rm c}}(\vec{k},\omega)$ at some $\vec{k}$ points near $\vec{k}=0$ (offset Gamma points).
The number of $\vec{k}$ points depends on the crystal symmetry.
${v}(\vec{k} \to 0)$ is replaced by $\overline{v}(\vec{k} \to 0)$, as well.
Then, we can take a simple discrete sum for both expressions of Eqs.~(\ref{eq:sigx}) and (\ref{sigmann}). 
The other point is how to perform an interpolation to give $\vxc_{\rm QSGW}$ at any $\vec{q}$ in the BZ.
For the interpolation, we expand the difference $\iDelta \vxc = \vxc_{\rm QSGW} - \vxc_{\rm LDA}$ in the MTOs, omitting APWs for expansion.
We have performed empty-sphere tests to complement the missing APWs \cite{Kotani2014}; however, including APWs, probably with the contracted plane wave \cite{Kotani2020CPW}, remains a subject for future work.

\section{Multi-GPU implementation of QSGW} \label{sec:multigpu}
Recently, multi-GPU computations have gained significant attention in scientific computing.
For example, GPUs are used for first-principles calculations in DFT, as described in Refs.~\cite{Jia2013, Das2022}.
In the $GW$ calculations, multi-GPU implementations have recently emerged \cite{Dimitar2020, Barker2020, Yu2022, Ness2024}.
One of the computationally expensive parts is the evaluation of the polarization functions. This part was successfully accelerated by GPUs in plane-wave basis codes. 
In QSGW, not only the polarization function but also the self-energy calculation are computationally expensive, since we have to calculate off-diagonal elements of the self-energy as in Eqs.~(\ref{eq:sigx}) and (\ref{sigmann}). Here, we demonstrate how to implement a multi-GPU version of QSGW in the \ecalj package.

\subsection{Computational cost of QSGW}
\begin{table*} [!tb]
  \caption{Size parameters in QSGW and computational cost of each part of the computation.
  To have an idea of the size of actual calculations, we show the size for a case of our benchmark for (InAs)$_{10}$(GaSb)$_{10}$ (40 atoms in a cell). See details in Table \ref{table:systemsize}.}
  \label{table:computationalcost}
  \begin{flushleft}
  \begin{tabular}{lll} \toprule
    Symbol    &  Description & (InAs)$_{10}$(GaSb)$_{10}$ \\ \midrule
    $\nIBZ$   &  number of $\vec{k}$ points in the irreducible Brillouin zone & 9\\
    $\nBZ$    &  number of $\vec{k}$ points in the Brillouin zone & 16 \\
    $\ncore$  & number of core states & 1180 \\
    $\nocc$   & number of occupied states & 460 \\
    $\nunocc$ & number of unoccupied states & 1348\\
    $\nall$   & number of all states =$\nocc$+$\nunocc$ & 1808 \\
    $\ncpb$   & number of product basis for core-exchange self-energy calculation & 8121\\
    $\npb$    & number of product basis & 5231 \\
    $\nmb$    & number of basis functions for eigenfunctions & 1834\\
    $\nqp$    & number of states to calculate self-energy & 713\\
    $\nspin$  & number of spin components & 1\\
    $\nomega$ & number of frequency mesh along real $\omega$ & 68 \\
    $\niomega$& number of imaginary frequency mesh (including $\omega=0$) & 11\\
    \bottomrule  
  \end{tabular}
  \vspace{20pt}
  \begin{tabular}{llllll} \toprule
                           & kernel   & Eq. & Computational cost           & Outer loop size& Array size\\ \midrule
    \multirow{2}{*}{\makecell{Core-exchange \\self-energy}}   & $Z$                    &  Eq. (\ref{eq:defz})    &  $\Order(\nqp \ncore \ncpb \nmb)$            & $\nIBZ \nBZ \nspin$ & \multirow{2}{*}{$(\nqp \ncore \ncpb)$}\\
                                     & $\varSigma^\mathrm{Cx}$&  Eq. (\ref{eq:sigx})      &  $\Order(\nqp^2 \ncore \ncpb)$               & $\nIBZ \nBZ \nspin$ &  \\ \midrule
    \multirow{2}{*}{\makecell{Exchange \\ self-energy} }        & $Z$                    &  Eq. (\ref{eq:defz})       &  $\Order(\nqp \nocc \npb \nmb)$              & $\nIBZ \nBZ \nspin$ & \multirow{2}{*}{$(\nqp \nocc \npb)$}\\
                                     & $\varSigma^\mathrm{x}$ &  Eq. (\ref{eq:sigx})       &  $\Order(\nqp^2 \nocc \npb )$                & $\nIBZ \nBZ \nspin$ &  \\ \midrule
    \multirow{4}{*}{\makecell{Screened \\ Coulomb \\ interaction}} & $Z$                    &  Eq. (\ref{eq:defz})       &  $\Order(\nocc \nocc \npb \nmb)$             & $\nIBZ \nBZ \nspin$ & \multirow{1}{*}{$(\nocc \nunocc \npb)$}\\ 
                                     & $\mathrm{Im}\varPi$    &  Eq. (\ref{eq:impolf})      &  $\Order(\nocc \nunocc \npb^2)$              & $\nIBZ \nBZ \nspin$ & \multirow{3}{*}{$(\npb^2 \nomega)$}\\
                                     & $\mathrm{Re}\varPi$    &  Eq. (\ref{eq:hilbert})  &  $\Order(\npb^2 \nomega)$                    & $\nIBZ$ & \\
                                     & $\varepsilon^{-1}v$    &  Eq. (\ref{eq:wc})       &  $\Order(\nomega \npb^3)      $              & $\nIBZ$ & \\ \midrule
    \multirow{3}{*}{\makecell{Correlation \\ self-energy}}    & $Z$                    &  Eq. (\ref{eq:defz})      &  $\Order(\nqp \nall \npb \nmb)$       & $\nIBZ \nBZ \nspin$   & \multirow{3}{*}{$(\nqp \nall \npb) $}\\
                                     & $R^\mathrm{c}$         &  Eq. (\ref{eq:sigcr})      &  $\Order(\nqp \nall \npb^2)$          & $\nIBZ \nBZ \nspin$ & \\
                                     & $I^\mathrm{c}$         &  Eq. (\ref{eq:sigci})      &  $\Order(\nqp \nall \npb^2 \niomega)$ & $\nIBZ \nBZ \nspin$ &  \\ \midrule
  \multirow{2}{*}{LDA}               & \multicolumn{2}{l}{Setup $H^0$} &  $\Order(\nmb^2 \nall)$     & $\nIBZ\nspin$ & \multirow{2}{*}{$(\nmb\nall)$}\\
                                     & \multicolumn{2}{l}{Solve $H^0\ket{\Psi_{\vec{k}n}} = \varepsilon_{\vec{k}n} \ket{\Psi_{\vec{k}n}}$} &  $\Order(\nmb^3)     $     & $\nIBZ\nspin$ & \\ \bottomrule
  \end{tabular}
\end{flushleft}
\end{table*}

QSGW calculations comprise four computationally intensive steps, as outlined in Sec.~\ref{sec:theory}.
Table \ref{table:computationalcost} shows
the order of computational costs for the computational kernels in each step, together with those in LDA.
To illustrate the size of actual calculations, we display the size for a benchmark, (InAs)$_{10}$(GaSb)$_{10}$. 
For supplied crystal structures, in addition to the number of divisions in the BZ, these numbers are determined by our default setting.
The key points regarding the sizes in Table \ref{table:computationalcost} are 
as follows:
(1) As shown in the numbers of (InAs)$_{10}$(GaSb)$_{10}$,
the number of product basis functions ($\npb$) is only several times 
larger than the number of basis for the wave function ($\nmb$).
In principle, we may expect $\npb/\nmb \sim 8= 2^3$ (to expand products of plane waves).
However, we had verified that such a large $\npb$ is not necessary in our construction of MPB \cite{Kotani2007}.
(2) We use the intermediate states $\nall = \nocc + \nunocc$ which
is almost the same as $\nmb$ (the number of basis functions $\simeq$ the eigenfunctions).
Thus, we handle the space spanned by the PMT basis functions very well, 
that the artificial truncation is avoided.
Some eigenfunctions are missing at the diagonalization step of $H^0$ because of the poor linear dependency of the overlap matrix.
(3) In our default setting, $\nqp$ (where $\nqp < \nall$) is determined 
to include quasiparticle states less than Fermi energy +2 $\sim$ +3 Ry.
We had checked that this is a suitable truncation to reduce computational costs \cite{Kotani2014}.

As shown in Table \ref{table:computationalcost}, the computational cost of $W$ and the correlation part of self energy are particularly high, making them the most dominant components of the QSGW calculation.
Based on our experience, the current computation time for QSGW
is approximately 500 times longer than that for LDA/GGA calculations in the 10-atom system because of the significantly higher computational cost compared to LDA/GGA calculations.
The fact that the size of the array scales with the cube of the size of the system indicates that memory partitioning is necessary for large systems.
The following section discusses the method for using GPUs in QSGW in \ecalj.

\subsection{Parallelization scheme with GPU}
\begin{figure}[tb]
  \centering
  \includegraphics[width=8.4cm]{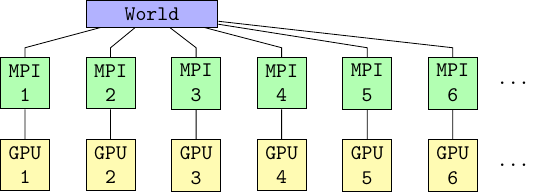}
  \caption{Parallelization scheme for QSGW calculations.}
  \label{fig:parallelization_hierarchy}
\end{figure}
We have two versions of code to perform QSGW, which are simply switched by a C preprocessor option (cpp).
One is just the MPI (CPU) version.
The other is the GPU version. In this case, we use one MPI process per GPU, as shown in Fig.~\ref{fig:parallelization_hierarchy}. 
In principle, our implementation enables the sharing of a GPU among multiple MPI processes by using Multiprocess-Service (MPS) or Multi-Instance-GPU (MIG) techniques \cite{Baolin2022}.
However, such an approach is only effective if the computational cost per MPI process is small and a single GPU resource cannot be fully utilized.
Additionally, it is unsuitable for large-scale systems because the GPU memory of each MPI process decreases.
Thus, we now arrive at the conclusion that the MPI process per GPU is a best practice, as in Fig.~\ref{fig:parallelization_hierarchy}.

Maintaining the computational code is vital for the continuation of future research and development.
It would be impractical if the GPU version were merely a one-off optimization that lacks maintainability.
Thus, we have carefully designed the above implementations.

At the MPI level, we optimize the Fortran codes. 
Subsequently, we introduce the GPU version by replacing the expensive parts with GPU coding.
Our code is based on module-based programming using the latest modern Fortran.
We utilize vector operations and minimize the use of do loops whenever possible.
Intrinsic functionalities such as \verb#block, findloc, counts, associate#, and array constructors are fully used.
In object-oriented programming terminology, we utilize modules to mimic the singleton pattern.
That is, modules contain various data generated simply by calling the subroutines in modules (initialization and/or some instructions).
The set of data can be read by other subroutines, but cannot be modified by them.
The Fortran singleton pattern is helpful for robust and efficient implementation.
One of the advantages is that one can safely read any data in any subroutine by directly accessing the modules where the data is stored.
It is essential to make no copies of the data.
For example, the number of atoms is defined as a module variable in a specific module. We access the number of atoms exclusively through this variable, which helps to avoid data inconsistency within the code.

In the GPU version, we utilize OpenACC programming to eliminate code duplication between the MPI and GPU versions as much as possible.
OpenACC enables the effective utilization of GPUs by inserting directives into existing code, allowing the source code to be shared between CPU and GPU versions.
Moreover, previous research has reported that GPU implementations of one-shot $GW$ calculation using OpenACC directives achieve performance comparable to or slightly lower than CUDA implementations \cite{Yu2022}.
It indicates that OpenACC is one of the suitable options for straightforwardly achieving high performance while ensuring high code maintainability.
However, it is important to note that implementing OpenACC programming alone has limitations. 
Due to its simplicity and limited functionality, OpenACC is more challenging to optimize and tune code than CUDA programming. 
Therefore, we employ a GPU-accelerated math library as much as possible for the kernel calculations. As mentioned, the $GW$ calculations involve tensor contractions, as shown in Eqs.~(\ref{eq:sigx}), (\ref{eq:impolf}), (\ref{eq:sigci}), and (\ref{eq:sigcr}), which can be executed through GPU-accelerated matrix multiplication.
Matrix multiplications are particularly well suited for GPUs; for example, recent NVIDIA GPUs have been able to accelerate matrix multiplication using Tensor Core.

Effective data transfer between the CPU and GPU is another crucial point in achieving high computational efficiency in GPU usage.
Since data transfer between CPU and GPU is quite time-consuming,
we design data control for GPU so that the expensive part of computations is performed within the GPU alone. Consequently, the CPU only performs small but complicated computations that are difficult to parallelize efficiently, as well as the data management and computational control.

\subsection{The GPU usage implementation procedure on \ecalj}
Generally, the development of GPU-accelerated computational code follows these steps:
\begin{enumerate}
  \item[I.] Identify the bottleneck parts of the computation
  \item[II.] Move the bottleneck computations to the GPU
  \item[III.] Optimize data transfer between CPU and GPU
\end{enumerate}
In Step I, performance analysis tools such as the Intel VTune Profiler are valuable tools.
Regarding Step II, we have developed low-level Fortran modules for core operations, such as matrix multiplication and matrix inversion calculations. 
These modules have CPU and GPU versions of core operations routines with a unified interface.
In the higher-level modules, simply switching the called routines allows the use of either CPU or GPU.
The called routines are switched using a combination of Fortran module aliases and preprocessor directives.
The data transfer and small-scale GPU operations in the high-level modules are controlled by OpenACC programming, ensuring common code for both CPU and GPU.
Furthermore, the variables used exclusively by the GPU are clearly defined using the `device` attribute in CUDA Fortran.
On the other hand, for public attribute variables used in both the CPU and GPU, the variable scope can become unclear, negatively affecting the code's maintainability.
Therefore, we restrict such variables to a limited number of large arrays.
Step III minimizes data transfer by offloading all large-scale data operations to the GPU.
During the development stage, we utilized NVIDIA Nsight Systems to identify the low-efficiency parts of GPU usage and the bottlenecks caused by data transfer.

To effectively implement GPU usage on \ecalj, we optimized our code before introducing GPU, focusing on the following two concepts.
First, it is crucial for reducing memory usage and enhancing the robustness of calculations, especially in large-scale systems.
We divided the internal $n_\mathrm{pb}$ in the calculation of $W$ using MPI, thereby reducing the memory usage per MPI process.
Moreover, segmented batch processing for the intermediate state of $Z$ is also implemented for these four steps because its size at $\vec{k}$ and $\vec{q}$ points is proportional to the cube of the system, which far exceeds the available memory in large-scale systems.
Second, as mentioned earlier, the kernel calculations in the \ecalj code were replaced with matrix multiplication.
To accomplish this, we restructured the arrays and changed the order of operations to utilize matrix operations.
With respect to this point, the following details the specific implementation of $W$ and the correlation self-energy $\varSigma$.

\subsubsection{Screened Coulomb interaction calculation}
\begin{algorithm*}[!tb]
  \caption{Pseudo code to calculate $W^{\rm c}$ with our MPI-GPU architecture shown in Fig.~\ref{fig:parallelization_hierarchy} .
  MPI communications are at lines \ref{line:mpi_reduce} and \ref{line:mpi_gather}. 
  At lines \ref{line:mpi1} and \ref{line:mpi2}, we assign a QKM block to the MPI process $p$. Exactly speaking,
  we need additional $\vec{q}$ and $\vec{k}$ for the offset-$\Gamma$ method evaluating $\overline{W^{\rm c}}(\vec{q}=0,\omega_j)$, 
  but not shown here explicitly. To calculate $Z$ at Line \ref{line:z}, we employ further MPI parallelization, which is not shown here.
  }
  \label{algo:hrcxq}
  \begin{algorithmic}[1]
  \State Bootstrap initialization to store data sets in modules. The data sets are for crystal structure, frequency mesh, 
  eigenvalues, eigenfunctions, and computational settings.
  \State Set up divisions for MPI processes as $\{\{\vec{q}\}_Q|Q=0,1,2,\cdots,N_Q-1\},\{\{\vec{k}\}_K|K=0,1,2,\cdots,N_K-1\},
  \{\{\mu\}_M|M=0,1,2,\cdots,N_M-1\}$. \label{line:mpi1}
  \State Generate functions $Q(p),K(p),M(p)$, so that a QKM  
  block $\{\vec{q}\}_{Q(p)} \times \{\vec{k}\}_{K(p)} \times  \{\mu\}_{M(p)}$ is assigned to the MPI process $p$,
  where $p=0,1,2,\cdots,N_\mathrm{MPI}-1$. \label{line:mpi2}
\State ----- Following steps are at each MPI process $p$ ----- 
  \For{$\vec{q} \gets \{\vec{q}\}_{Q(p)}$}
    \State {$\mathrm{Im}\varPi_\vec{q}(1\!:\!n_\text{pb},\{\mu\}_{M(p)},1\!:\!n_\omega) \gets 0$} \Comment{$\varPi$ for $\vec{q}$ will be stored on $\varPi_\vec{q}$}
    \State {Read $v_\vec{q}(1\!:\!n_{pb})$} \Comment{$v_\mu(\vec{q})$ in Eq.~(\ref{eqvcoue})}
    \State {Calculate tetrahedron weights $w(1\!:\!n_{\rm unocc},1\!:\!n_{\rm occ},1\!:\!n_\omega,\{\vec{k}\})$} \Comment{$w_{n'n}^{\vec{k},\vec{q}}$ in Eq.~(\ref{eq:tetw})}
    \For{$\vec{k} \gets \{\vec{k} \}_{K(p)}$} 
      \State Batch division $1\!:\!n_{\rm unocc}$ into $\{n'\}_b$ where $b=1,2,\cdots,B^\mathrm{batch}$.\label{line:x0batch} \Comment{Avoid huge memory for $Z$} 
      \For{$b \gets 1\!:\!B^\text{batch}$}
        \State {Calculate $Z(1\!:\!n_\text{pb},\{n'\}_b,1\!:\!n_\text{occ})$ } \Comment{For ${\vec{k},\vec{q}}$ in Eq.~(\ref{eq:defz})} \label{line:z}
        \For{$i_\omega \gets 1\!:\!n_{\omega}$} \Comment{$i_\omega$ specifies the histogram bins to accumulate the imaginary part}
          \State Set up pair index $\{n'(i_\mathrm{pair}),n(i_\mathrm{pair})| 
          i_\mathrm{pair}=1,2,\cdots,n_\mathrm{pair}\}$ 
          for $w(n'(i_\mathrm{pair}),n(i_\mathrm{pair}),i_\omega,\vec{k}) \neq 0$ and $n'(i_\mathrm{pair}) \in \{n'\}_b$
          \State \texttt{!\$acc kernel}
          \For{$i_\mathrm{pair} \gets 1\!:\!n_\mathrm{pair}$}
            \State $Z_\mathrm{pair} (i_\mathrm{pair},1\!:\!\npb) \gets Z(1\!:\!\npb, n'(i_\mathrm{pair}), n(i_\mathrm{pair}))$
            \State $W_{Z_\mathrm{pair}} (i_\mathrm{pair},\{\mu\}_{M(p)}) \gets Z_\mathrm{pair} (i_\mathrm{pair},\{\mu\}_{M(p)}) 
            w(n'(i_\mathrm{pair}),n(i_\mathrm{pair}),i_\omega,\vec{k})$
          \EndFor
          \State \texttt{!\$acc end kernel}
          \State $\mathrm{Im}\varPi_\vec{q}(1\!:\!n_\text{pb},\{\mu\}_{M(p)},i_\omega) \gets \mathrm{Im}\varPi_\vec{q}(1\!:\!n_\text{pb},\{\mu\}_{M(p)},i_\omega) 
          + Z_\mathrm{pair}^\dagger \times W_{Z_\mathrm{pair}}$  \Comment{GEMM}
        \EndFor
      \EndFor
    \EndFor
    \If{$K(p)==0$}  \Comment{Reduction for $\vec{k}$}
     \State{$\varPi_\vec{q}(1\!:\!n_\text{pb},\{\mu\}_{M(p)},i_\omega) 
     \gets \sum_{p'}^{\vec{q} \in Q(p'),M(p')==M(p)} \varPi^{p'}_\vec{q}(1\!:\!n_\text{pb},\{\mu\}_{M(p)},i_\omega)$} \label{line:mpi_reduce}
     \Comment{MPI reduce sum. Superscript $p'$ of $\varPi^{p'}_\vec{q}$ indicates data belonging to the process $p'$.}
     \State {Perform Hilbert transformation to $\mathrm{Im}\varPi_\vec{q}$, resulting in
     $\varPi_\vec{q}$ at $\{\omega_j| \omega_j \text{ along Real and Imag axes}\}$}. \Comment{GEMM}
     \For {$j_\omega$ for $\{\omega_j| \omega_j \text{ along Real and Imag axes}\}$} 
     \If{$M(p)==0$}  \Comment{Reduction for MPB}
     \State {$\varPi_{\vec{q}{j_\omega}}(1\!:\!n_\text{pb},1\!:\!n_\text{pb})
       \gets \bigoplus_{p'} \varPi^{p'}_\vec{q}(1\!:\!n_\text{pb}, \{\mu\}_{M(p')},j_\omega)$} \label{line:mpi_gather}
      \Comment{MPI gather. Superscript $p'$ of $\varPi^{p'}_\vec{q}$ indicates data belonging to the process $p'$.}
     \State {Calculate $\epsilon(1\!:\!n_\text{pb},1\!:\!n_\text{pb})$ from $v_{\vec{q}}$ and $\varPi_{\vec{q}{j_\omega}}$.} \Comment{For $\vec{q},\omega=\omega_j$ in Eq.~(\ref{eq:dielectricfunction}), OpenACC}
      \State {Calculate $\epsilon^{-1}(1\!:\!n_\text{pb},1\!:\!n_\text{pb})$} \Comment{For $\vec{q},\omega=\omega_j$ in Eq.~(\ref{eq:wc}), MATINV}
      \State {Calculate $W^\mathrm{c}(1\!:\!n_\text{pb},1\!:\!n_\text{pb})$} \Comment{For $\vec{q},\omega=\omega_j$ in Eq.~(\ref{eq:wc}), OpenACC }
    \State {Write $W^\mathrm{c}(1\!:\!n_\text{pb},1\!:\!n_\text{pb})$ into a file}. \Comment{At $\vec{q}=0$, $\overline{W^\mathrm{c}}(1\!:\!n_\text{pb},1\!:\!n_\text{pb})$ instead.}
      \EndIf
      \EndFor
    \EndIf
  \EndFor
  \end{algorithmic}
\end{algorithm*}

Due to the above development, our QSGW calculations code works with massive MPI parallelization, regardless of GPU usage.
In the calculation of $W$, computations for specific $\vec{q}$ points are executed within each divided MPI communicator.
Within the intracommunicator, MPI parallelization of MPB and $\vec{k}$ points is implemented.
In our implementation, $\varPi$ and $W$ are the largest arrays, whose specific size will be described later; therefore, this memory partitioning is crucial for practical applications.
MPB parallelization enables the memory partitioning for $\varPi$, thereby reducing the array size of $\varPi$ stored in each MPI process.
For this reason, MPB parallelization is prioritized over $\vec{k}$-point parallelization; therefore, the latter is set to one in our QSGW calculations. 
Meanwhile, $\vec{k}$-point parallelization can be utilized for the calculation of the macroscopic dielectric function in the further analysis stage because it usually requires dense $\vec{k}$-point sampling to reach convergence \cite{Gajdos2006}.
When calculating the dielectric function ($\epsilon$) from $\varPi$, as shown in Eq.~(\ref{eq:dielectricfunction}), the data within the intra-communicator about MPB is gathered.
The inverse matrix in Eq.~(\ref{eq:screenedcoulomb}) is obtained using cuSOLVER in the GPU version.

The kernel part of the $\varPi$ calculation involves constructing $Z$ in Eq.~(\ref{eq:defz}) and sum for the occupied orbitals $n$ and for the unoccupied orbitals $n'$ at $\vec{k}$ points, described in Eq.~(\ref{eq:impolf}).
The overall implementation is illustrated in Alg. \ref{algo:hrcxq}, where $\{\vec{q}\}_Q, \{\vec{k}\}_K$ and $\{\mu\}_M$ indicate $\vec{q}$-point, $\vec{k}$-point, and MPB parallelization, respectively.
GEMM (general matrix multiplication) and OpenACC directives (\texttt{\$acc}) indicate the regions to be computed on the GPU; the detailed specifications are omitted for simplicity.
As mentioned earlier, segmented batch processing of intermediate states, due to the array size of $Z$ exceeding the available memory, is introduced as an internal loop, as described in Alg. \ref{algo:hrcxq}, line \ref{line:x0batch}.
As shown in Eq.~(\ref{eq:tetrahedronweeight}), $w$ only has contributions from specific $n$ and $n'$ for a given $\omega$.
Reconstructing $w$ and $Z$ only from $n$ and $n'$ that contribute to the summation, we can perform the summation using the GEMMs routines.
The calculations of $Z$ and the summation part of $\varPi$, which are the predominant components of $\varPi$ computation, as well as all computations involving $Z$ and $\varPi$, are performed on the GPU.
Therefore, the primary data transfers between the CPU and GPU involve reading the wave functions, which are the results of \texttt{lmf}, from files and transferring them to the GPU, as well as copying $W$ created on the GPU to the CPU for file output.

\subsubsection{Static version of correlation self-energy calculation}

\begin{algorithm*}[htbp]
  \caption{Pseudo code to calculate ${\varSigma}^\mathrm{c}$ (\texttt{hsfp0\_sc}) with our MPI-GPU architecture. MPI communications are at line \ref{line:mpi_reduce_sc}. 
  This is for the correlation self-energy. Ensure that frequency $\omega_i$ and weight $w_i$ are not confused.} 
\label{algo:hsfp0}
\begin{algorithmic}[1]
  \State Bootstrap initialization to store data sets in modules. Data sets are for crystal structure, frequency mesh, 
  eigenvalues, eigenfunctions, and computational settings.
  \State Set up divisions for MPI processes as 
  $\{\{\vec{q}\}_Q|Q=0,1,2,\cdots,N_Q-1\},\{\{\vec{k}\}_K|K=0,1,2,\cdots,N_K-1\}$,
  $\{\{\omega^\mathrm{I}_i\}_{\Omega}|\Omega=0,1,2,\cdots,N^\Omega-1\},
  \text{ as well as }\{\{\omega^\mathrm{R}_i\}_\Omega|\Omega=0,1,2,\cdots,N^\Omega-1\}$.
 \label{line:mpi_sc1}
\State Generate functions $Q(p),K(p),\Omega(p)$ for assigning a QK$\Omega^\mathrm{I}$  
block $\{\vec{q}\}_{Q(p)} \times \{\vec{k}\}_{K(p)} \times \{\omega^\mathrm{I}_i\}_{\Omega(p)}$  
to the MPI process $p$, where $p=0,1,2,\cdots,N_\mathrm{MPI}-1$. 
  In the same manner, $Q(p),K(p),\Omega^\mathrm{R}(p)$ for a QK$\Omega^\mathrm{R}$ as well. \label{line:mpi_sc2}
  \State ----- Following steps are at each MPI process $p$ -----
  \State ${\varSigma}^\mathrm{c}(1\!:\!n_\text{qp},1\!:\!n_\text{qp},\{\vec{q}\}) \gets 0$ \Comment{For $\varSigma^\text{c}(\vec{q},\varepsilon_{\vec{q}n})$}
  \For{$\vec{q} \gets \{\vec{q}\}_{Q(p)}$}
    \State {Read $v_\vec{q}(1\!:\!n_{pb})$} \Comment{$v_\mu(\vec{q})$ in Eq.~(\ref{eqvcoue})}
    \Ifoneline{GPU version}{Read 
      $W^\mathrm{c}_\mathrm{I}(1\!:\!n_\text{pb},1\!:\!n_\text{pb},\{\omega^\mathrm{I}\}_{\Omega(p)})$,
      $W^\mathrm{c}_\mathrm{R}(1\!:\!n_\text{pb},1\!:\!n_\text{pb},\{\omega^\mathrm{R}\}_{\Omega(p)})$} 
      \Comment{$W^\mathrm{c}(\vec{q},\omega)$ in Eqs.~(\ref{eq:sigci}) and (\ref{eq:sigcr})}
    \For {$\vec{k} \gets \{\vec{k}\}_{K(p)}$}
      \State Batch division $1\!:\!n_{\rm all}$ into $\{n'\}_b$ where $b=1,2,\cdots,B^\mathrm{batch}$. \Comment{Avoid huge memory for $Z$} 
      \For{$b \gets 1\!:\!B^\text{batch}$}
      \State $ZW^\mathrm{c}(\{n'\}_b,1\!:\!n_\mathrm{qp},1\!:\!n_\mathrm{pb}) \gets 0$
      \State {Calculate $Z(1\!:\!n_\text{pb},\{n'\}_b,1\!:\!n_\text{qp})$ } \Comment{For ${\vec{k},\vec{q}}$ in Eq.~(\ref{eq:defz})} \label{line:z2}

      \State { ----- QK$\Omega^\mathrm{I}$ block -----}
        \State {Calculate the weight 
        $\{w^\mathrm{I}_i(\varepsilon_{\vec{q} n}-\varepsilon_{\vec{q}-\vec{k} n'})|n' \in \{n'\}_b, 
        n \in n_\mathrm{qp}, \omega_i^\text{I} \in \{\omega_i^\text{I}\}_{\Omega(p)}\}$} 
        \Comment{Eq.~(\ref{eq:sigci})}
        \For {$\omega_i^\mathrm{I} \gets \{\omega_i^\mathrm{I}\}_{\Omega(p)}$} \Comment{QK$\Omega^\mathrm{I}$ block}  \label{line:qkoi}
          \State \texttt{!\$acc kernel}
          \For{$n \gets 1\!:\!n_\text{qp}$, $n' \in $ $\{n'\}_{b}$}
          \State $Z_w (1\!:\!\npb,n', n) \gets Z(1\!:\!\npb, n',n) 
          w^\mathrm{I}_i(\varepsilon_{\vec{q} n}-\varepsilon_{\vec{q}-\vec{k} n'})$
          \EndFor
          \State \texttt{!\$acc end kernel}
          \Ifoneline{GPU version}{$W^\mathrm{c}(1\!:\!n_\text{pb}, 1\!:\!n_\text{pb}) \gets W^\mathrm{c}_\mathrm{I}(1\!:\!n_\text{pb},1\!:\!n_\text{pb},\omega^\mathrm{I})$}
          \Ifoneline{CPU version}{Read $W^\mathrm{c}(1\!:\!n_\text{pb}, 1\!:\!n_\text{pb})$ from a file.}
          \Comment{For $\vec{q}, \omega^\mathrm{I}$ of $W^\mathrm{c}(\vec{q}, i\omega^\text{I})$ in Eq.~(\ref{eq:sigci})}
          \State $ZW^\mathrm{c} \gets ZW^\mathrm{c} + Z_w^\dagger \times W^{c}$ \Comment{GEMM, accumulate the first dimension of $Z_w$ and $W^\mathrm{c}$.}
        \EndFor
        \State {Reshape $ZW^\text{c}(\{n'\}_b,1\!:\!n_\mathrm{qp},1\!:\!n_\mathrm{pb})$ to $ZW^\text{c}(1\!:\!n_\text{pb},\{n'\}_b,1\!:\!n_\text{qp})$} \Comment{OpenACC}

        \State { ----- QK$\Omega^\mathrm{R}$ block -----}
        \State {
          Calculate the weight 
          $\{w_i^\mathrm{R}(\varepsilon_{\vec{q} n}-E_{\rm F},\varepsilon_{\vec{q}-\vec{k} n'}-E_{\rm F})
          |n' \in \{n'\}_b, n \in n_\mathrm{qp}, \omega^R_i \in \{\omega^\mathrm{R}_i\}_{\Omega(p)}\}$}  
          \Comment{Eq.~(\ref{eq:sigcr})}
        \State Set up pair index $\{n_i'(j_\mathrm{pair}),n_i(j_\mathrm{pair})| 
        j_\mathrm{pair}=1,2,\cdots,n_{\mathrm{pair},i}\}$ 
          for $w_i^\mathrm{R}(\varepsilon_{\vec{q} n_i(j_\mathrm{pair})}-E_{\rm F},\varepsilon_{\vec{q}-\vec{k} n'_i(j_\mathrm{pair})}-E_{\rm F}) \neq 0$, 
          $\omega^R_i \in \{\omega^\mathrm{R}_i\}_{\Omega(p)}$, and $n'_i(j_\mathrm{pair}) \in \{n'\}_b$
  
          \For {$\omega^\mathrm{R}_i \gets \{\omega^\mathrm{R}_i\}_{\Omega(p)}$}  \Comment{QK$\Omega^\mathrm{R}$ block} \label{line:qkor}
          \State \texttt{!\$acc kernel}
          \For{$j_\mathrm{pair} \gets 1\!:\!n_{\mathrm{pair},i}$}
          \State $Z_w (1\!:\!\npb,j_\mathrm{pair}) \gets Z(1\!:\!\npb,n_i'(j_\mathrm{pair}),n_i(j_\mathrm{pair}))
          w_i^\mathrm{R}(\varepsilon_{\vec{q} n_i(j_\mathrm{pair})}-E_{\rm F},\varepsilon_{\vec{q}-\vec{k} n_i'(j_\mathrm{pair})}-E_{\rm F})$
          \EndFor
          \State \texttt{!\$acc end kernel}
          \Ifoneline{GPU version}{$W^\mathrm{c}(1\!:\!n_\text{pb}, 1\!:\!n_\text{pb}) \gets W^\mathrm{c}_\mathrm{R}(1\!:\!n_\text{pb},1\!:\!n_\text{pb},\omega_i^\mathrm{R})$}
          \Ifoneline{CPU version}{Read $W^\mathrm{c}(1\!:\!n_\text{pb}, 1\!:\!n_\text{pb})$ from a file.}
          \Comment{For $\vec{q}, \omega^\mathrm{R}$ of $W^\mathrm{c}(\vec{q}, \omega)$ in Eq.~(\ref{eq:sigcr})}
          \State $Z_wW^\mathrm{c} \gets Z_{w}^\dagger  \times  [W^{\mathrm{c}} + W^{\mathrm{c\dagger}}]/2$ \Comment{GEMM}
          \State \texttt{!\$acc kernel}
          \For{$j_\mathrm{pair} \gets 1\!:\!n_{\mathrm{pair},i}$}
          \State $ZW^\mathrm{c}(1\!:\!\npb,n_i'(j_\mathrm{pair}),n_i(j_\mathrm{pair})) \gets ZW^\mathrm{c}(1\!:\!\npb,n'_i(j_\mathrm{pair}),n_i(j_\mathrm{pair}))
          + Z_wW^\mathrm{c}(j_\mathrm{pair}, 1\!:\!\npb)$
          \EndFor
          \State \texttt{!\$acc end kernel}
        \EndFor
        \State ${\varSigma}^\mathrm{c}(1\!:\!n_\mathrm{qp},1\!:\!n_\mathrm{qp},\vec{q}) 
        \gets {\varSigma}^\mathrm{c} (1\!:\!n_\mathrm{qp},1\!:\!n_\mathrm{qp},\vec{q}) 
        + \smashoperator{\sum_{i_\mathrm{PB} \in n_\mathrm{PB},n' \in \{n'\}_b}} ZW^\mathrm{c}(i_\mathrm{PB},n',1\!:\!n_\mathrm{qp}) \times Z(i_\mathrm{PB},n',1\!:\!n_\mathrm{qp})$ 
        \Comment{GEMM} \label{line:zwz}
      \EndFor
    \EndFor
  \EndFor
  \If{$K(p)==0$ and $\Omega(p)==0$ }  \Comment{Reduction for $\vec{q}, \vec{k}, \omega$}
  \State {${\varSigma^\mathrm{c}}(1\!:\!n_\text{qp},1\!:\!n_\text{qp}, \{\vec{q}\})
  \gets \sum_{p'} {\varSigma}^{\mathrm{c}p'}(1\!:\!n_\text{qp},1\!:\!n_\text{qp}, \{\vec{q}\})$} \label{line:mpi_reduce_sc}
  \Comment{${\varSigma}^{\mathrm{c}p'}$ denotes ${\varSigma}^{\mathrm{c}}$ at he MPI process $p'$. MPI reduce sum.} 
  \State {Write $\varSigma^{\mathrm{c}}(1\!:\!n_\text{qp},1\!:\!n_\text{qp}, \{\vec{q}\})$}
  \EndIf
\end{algorithmic}
\end{algorithm*}

In the implementation of the static version of correlation self energy, the \texttt{World} is divided by MPI parallelization for $\vec{k}$, $\vec{q}$, and $\omega'$.
The kernel operations involve the construction of $Z$ and the $\omega'$ integration shown in Eqs.~(\ref{eq:sigci}) and (\ref{eq:sigcr}), as well as the summation over $n'$ and MPB.
Algorithm \ref{algo:hsfp0} outlines the pseudo code for the calculation of $\widetilde{\varSigma}^\mathrm{c}$.
The GEMM operations are repeated for $\omega_i^\mathrm{I}$ to perform the imaginary axis integration, which is the most time-consuming calculation in the correlation self energy.
The real axis integration only has values between specific orbitals ($n$, $n'$); therefore, a similar implementation can be employed for the $\mathrm{Im}\varPi$.
Note that since this is the calculation of the static version of self energy, the real-axis integration involves the hermitization of $W$, leading to the vanishing of the imaginary component of self energy.
Although computation can be accelerated on the GPU, file reading, which involves data transfer via the CPU, becomes a significant bottleneck in the GPU version.
Therefore, the $W$ data is stored in memory in advance on the GPU version. 
In contrast, it is read from the file each time on the CPU version because this treatment is not applicable to the CPU version due to the insufficient memory of each MPI process.

The MPI parallelization for $\omega'$ is less efficient than that of $\vec{k}$ and $\vec{q}$ due to the calculation of $Z$ outside of the $\omega'$ loop.
Therefore, the division of $\vec{k}$ and $\vec{q}$ is basically prioritized over $\omega'$ parallelization.
However, the division of $\omega'$ has the advantage of reducing the array size of $W$ to be stored in memory, which helps avoid GPU memory exhaustion in the GPU version where $W$ is stored in GPU memory.
Consequently, the $\omega'$ parallelization is used in the largest system of this benchmark (InAs)$_{10}$(GaSb)$_{10}$ in the GPU version, but it is inactive in other cases.

\subsection{Mixed precision approach}
In GPU utilization, using single-precision or Tensor-Float (TF) operations, such as TF32, can improve computational speed at the expense of computational accuracy.
For example, on NVIDIA's A100 GPU, matrix multiplication using TF32 operations has eight times the computational performance compared to double precision (DP) with Tensor Core operations.
We have confirmed that using TF32 operations in complex matrix multiplication benchmarks improves computational speed by more than 5 times.
Since our computation-heavy tasks of the screened Coulomb interaction and self-energy calculations are mainly performed by complex matrix multiplication, the mixed precision (MP) approach is expected to increase computational speed drastically.
Moreover, the array size will be half that of the DP case, which is beneficial for memory usage.
Based on such motivations, we have implemented MP operations in GPU calculations.
However, since TF32 operations reduce computational precision, they can affect the accuracy of the results.
Thus, a verification of the computational accuracy of MP is a necessity for the application.
Note that our MP implementation does not replace all GPU operations with single-precision or TF32.
For example, operations such as matrix inversion and the calculation of the IPW overlap matrix, which may require high numerical precision, are performed using DP arithmetic.

The numerical accuracy of the MP approach may depend on the materials and computational settings.
While further validation across a broader range of materials and computational settings is necessary, this study primarily focuses on two benchmark systems to evaluate the computational efficiency and numerical accuracy of the MP approach.

\section{Benchmark and Discussion} \label{sec:benchmark}

\subsection{Benchmark system}
\begin{figure*}[!tb]
  
\centering
  \includegraphics[width=0.8\linewidth]{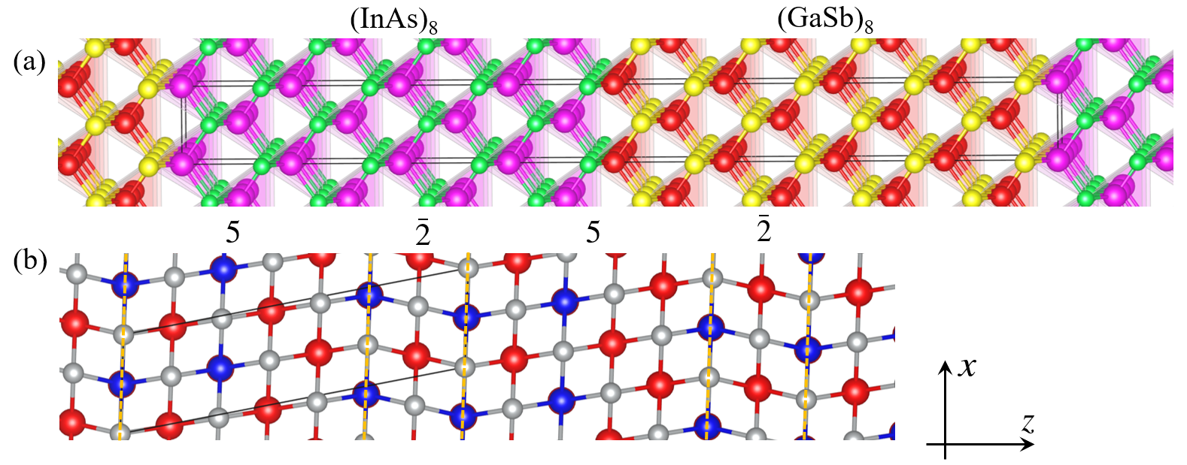}
\caption{
Structures of (InAs)$_8$(GaSb)$_8$(a) and Ni$_2$MnGa 14M (b), where In, As, Ga, Sb, Mn, and Ni are represented by pink, green, red, yellow, blue, and gray spheres, respectively.
The solid black lines represent the unit cell, and the dashed orange lines in (b) represent the boundaries of the nano twins.
}
  \label{fig:str}
\end{figure*}
\begin{table*}[tb]
  \centering
  \caption{
    Setting parameters, where these values represent the maximum in all $\vec{k}$ points. $n_\mathrm{val}$ is the number of valence electrons, which matches $n_\mathrm{occ}$ in insulator systems. $n_\mathrm{atom}$ is the number of atoms in the unit cell.
    The array size of $Z$ ($Z_\mathrm{size}$) is per $\vec{k}$ and $\vec{q}$ point and size of $W$ ($W_\mathrm{size}$) is per $\vec{q}$ point.
    These sizes become half in the case of the MP approach.
    In (InAs)$_n$(GaSb)$_n$, $n_\mathrm{BZ}$ of 16 (32) came from $4\times 4 \times 1$ ($4\times 4\times 2$), while in Ni$_2$MnGa, $n_\mathrm{BZ}$ of 24 (48) corresponds to $4\times 3\times 2$ ($4\times 3 \times 4$).
    The numbers in parentheses at $n_\mathrm{BZ}$ and $n_\mathrm{IBZ}$ indicate the values used in calculating the band gap.
  }
  \label{table:systemsize}
  \begin{tabular}{c|rrrrrcrrr} \toprule
                    & \multicolumn{5}{c}{(InAs)$_{n}$(GaSb)$_n$} & & \multicolumn{3}{c}{Ni$_2$MnGa} \\
                    & $n=2$ & $n=4$  & $n=6$ & $n=8$ & $n=10$ & & NM & 10M & 14M \\ \midrule
  $n_\mathrm{atom}$       & 8     & 16     & 24   & 32   & 40    & & 8   & 20  & 28 \\
  $n_\mathrm{spin}$       & 1     & 1      & 1    & 1    & 1     & & 2   & 2   & 2  \\
  $n_\mathrm{core}$       & 236   & 472    & 708  & 944  & 1180  & & 132 & 330 & 462\\
  $n_\mathrm{val}$        & 92    & 184    & 276  & 368  & 460   & & 92  & 230 & 322\\
  $n_\mathrm{qp}$         & 206   & 287    & 428  & 571  & 713   & & 123 & 301 & 421\\
  $n_\mathrm{all}$        & 361   & 722    & 1090 & 1444 & 1808  & & 269 & 670 & 940\\
  $n_\mathrm{mb}$         & 366   & 732    & 1106 & 1464 & 1834  & & 281 & 690 & 971\\
  $n_\mathrm{pb}$         & 1356  & 2097   & 3139 & 4189 & 5231  & & 982 & 2479& 3471\\
  $n_\mathrm{Cpb}$        & 1934  & 3253   & 4873 & 6501 & 8121  & & 1284& 3199& 4479\\
  $n_\mathrm{BZ}$         & 16(32)& 16(32) & 16   & 16   & 16    & & 48  & 24  & 24 \\
  $n_\mathrm{IBZ}$        & 9(18) & 9(18)  &  9   & 9    & 9     & & 14  & 14  & 14 \\
  $Z_\mathrm{size}$ (GiB) & 1.50  & 6.47   & 21.82 & 51.47& 100.48 & & 0.48 & 7.45 & 20.47 \\
  $W_\mathrm{size}$ (GiB) & 2.22  & 5.31  &  11.89 & 21.18 & 33.03 & & 1.16 & 7.42 & 14.54 \\ \bottomrule
  \end{tabular}
\end{table*}
In this study, we benchmark two physical systems: a Type-II superlattice of the semiconductor compound (InAs)$_n$(GaSb)$_n$ $(n=2, 4, 6, 8, 10)$ and the modulated martensite (10M, 14M) and non-modulated (NM) phases of Ni$_2$MnGa.
The former has a narrow direct band gap that can be adjusted by the number of stacking layers ($n$), making it suitable for infrared sensor applications \cite{Michel2011}.
The latter is known as a ferromagnetic shape memory alloy and is expected to be applied to functional materials, such as high-speed actuators \cite{Ullakko1996, Heczko2013, Sozinov2013}.
In both cases, issues with the physical accuracy of the evaluation of electronic correlations have been pointed out in calculations at the LDA/GGA level, and therefore analysis using more precise methods such as QSGW is desired \cite{Otsuka2017, Baigutlin2020, Obata2023}.

Figure \ref{fig:str} illustrates their crystal structures.
In (InAs)$_n$(GaSb)$_n$, the unit cells are tetragonal, and the construction of the structures follows the data described in Ref.~\cite{Otsuka2017}.
We assumed a constant interface distance between InAs and GaSb.
In Ni$_2$MnGa, the unit cells of the modulated phases are triclinic, and the structure is reproduced by building tetragonal blocks, where 10M (14M) has a 3$\bar{2}$3$\bar{2}$ (5$\bar 2$5$\bar 2$) shearing sequence.
The aspect ratio of the Ni tetragonal sublattice ($c/a$) and the density are determined from experimental values, specifically $c/a=1.21$ and \SI{49.4115}{\angstromcubic/f.u.}. \cite{Webster1984}.
The scale of the calculations is shown in Table \ref{table:systemsize}.
As mentioned above, since $Z_\mathrm{size}$ and $W_\mathrm{size}$ exceed the memory capacity available to a single MPI, they are handled in segments.
The segment size is determined on the basis of the available memory size.
Note that, in general, the amount of memory available per MPI process differs between the CPU and GPU versions, resulting in different segmented batch sizes for intermediate states.

All benchmarks are performed at the Supercomputer Center, ISSP, Univ. of Tokyo (system C: Kugui).
For the GPU version, the ACC one node in Kugui installed a CPU (AMD EPYC 7763), and four GPUs (NVIDIA A100 40GB for HGX) are used with four MPI processors.
For the CPU version, four CPU nodes in Kugui, with a total of 8 CPUs (AMD EPYC 7763), are used with 128 MPI with four-thread parallelization.
As shown in Tables \ref{table:computationalcost} and \ref{table:systemsize}, the QSGW calculations use significantly more memory than the LDA calculations.
Although memory segmentation is implemented, this parallel setting ensures that each MPI process has the necessary memory to perform the calculations.
This thread parallelization is achieved only through the MKL BLAS and LAPACK libraries.
The current efficiency of thread parallelization (4 threads) is measured to be 1.6 times faster than the flat MPI version in the case of (InAs)$_4$(GaSb)$_4$.
The observed reduction in thread parallelization efficiency can be attributed to the decreased efficiency of matrix operations.
This phenomenon has been observed in our benchmarks of complex matrix multiplication: suppression of thread-parallel efficiency occurs in highly MPI-parallelized computations.
The NVIDIA HPC SDK Version 24.7 compiler and its GPU-accelerated math libraries (cuBLAS, cuSOLVER) are used for our implementation.

\subsection{Benchmark results}
\begin{figure}[tb]
  \centering
  \includegraphics[width=8cm]{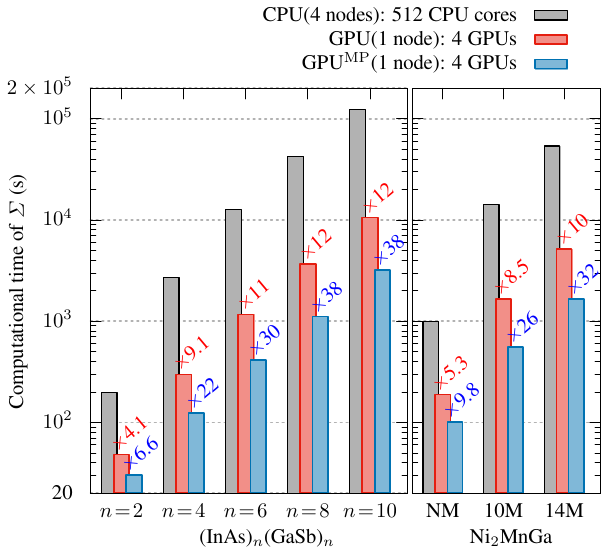}
  \caption{Computational time of $\varSigma$.
The meaning of the key is as follows. CPU (4 nodes): This is the reference to measure the speed of GPU computations. The total number of cores is 128$\times$4. 
GPU (1 node): Use one node with four GPUs. 
GPU$^\mathrm{MP}$ (1 node): Use one node with four GPUs with mixed precision GPU code.
In the figure, $\times 12$ means that the computational speed of the GPU (1 node) is worth (128 $\times$ 4) $\times 12$ cores (48 nodes).
Note that the CPU (4 nodes) results were already below the linear scaling proportional to the number of cores, despite our efforts to use the CPU cores as efficiently as possible. See the text for details.
 }
  \label{fig:etime_all}
\end{figure}
\begin{figure*}[!tb]
  \centering
  \includegraphics[width=0.90\linewidth]{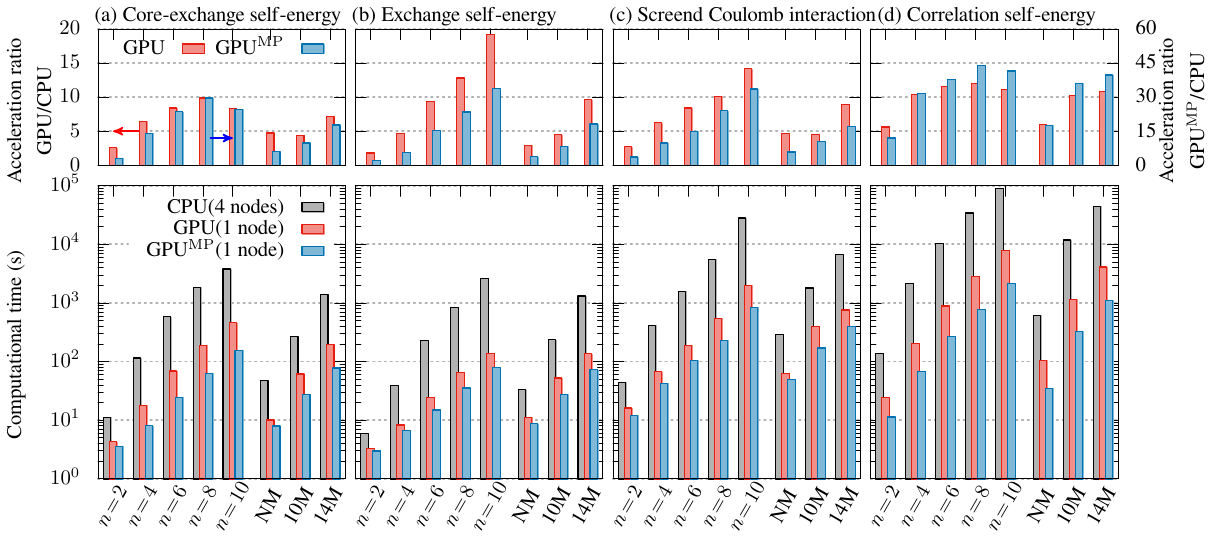}
  \caption{Computational time of the predominant four tasks. For the acceleration ratio, see the left (right) vertical axis for  between GPU (GPU$^\mathrm{MP}$) and CPU.}
  \label{fig:etime_detail}
\end{figure*}
\begin{figure*}[!tb]
  \centering
  \includegraphics[width=0.90\linewidth]{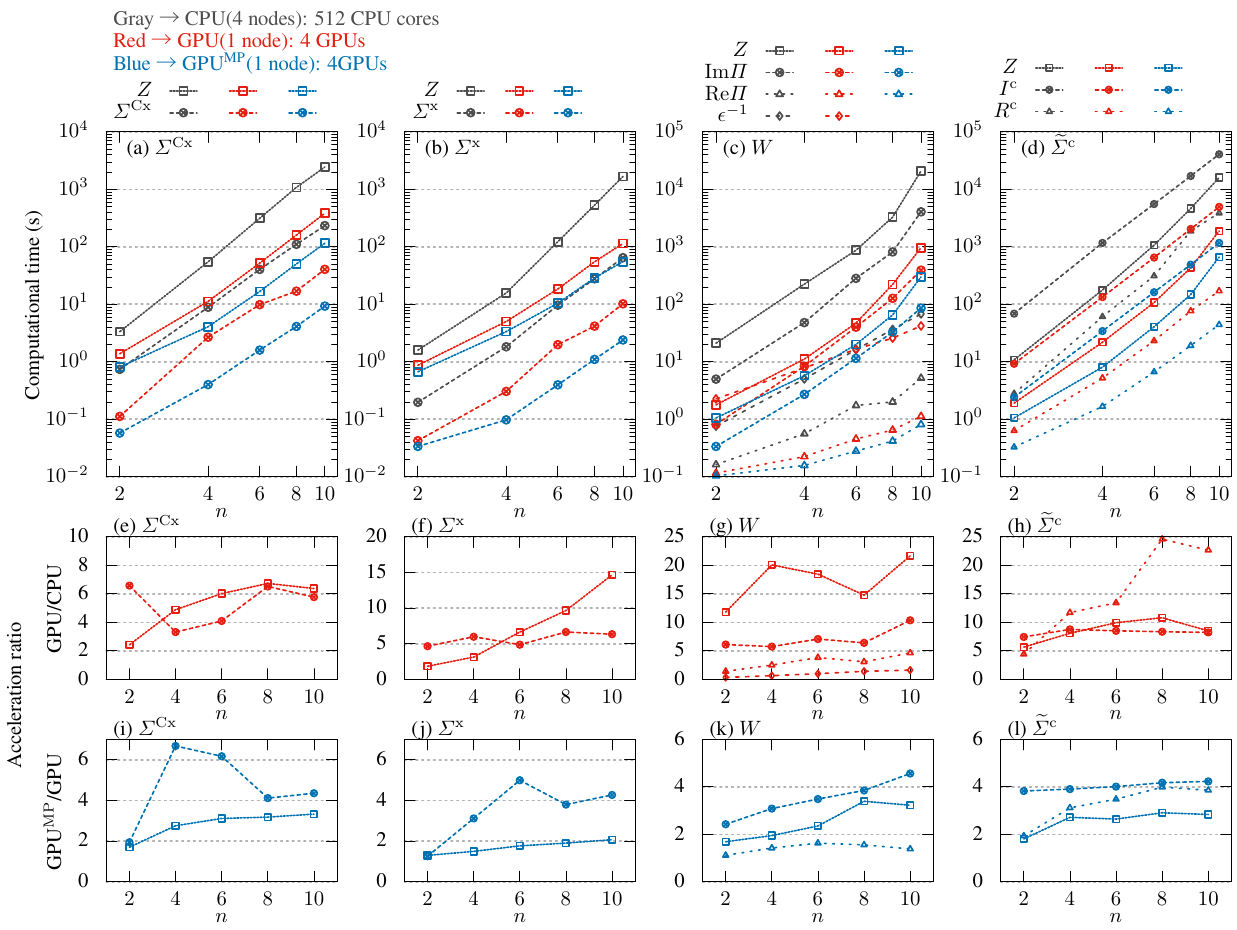}
  \caption{Computational time of kernel calculation in the predominant four tasks and those of acceleration ratio between GPU (GPU$^\mathrm{MP}$) and CPU.
  In the topmost figures, the vertical and horizontal axes are set to a logarithmic scale.}
  \label{fig:ctime}
\end{figure*}

Figure \ref{fig:etime_all} shows the total computational time (elapsed time) for the four main tasks in one cycle of QSGW calculations.
This comparison examines the calculation time performed on four CPU nodes versus one GPU node.
The computational time on the GPU is successfully reduced compared to that of the CPU.
In the paper, we define the acceleration ratio as the ratio of computational time between different approaches.
For example, the acceleration ratio of GPU/CPU is estimated from the computational time of the CPU divided by that of the GPU.
In (InAs)$_n$(GaSb)$_n$ with ( $n \geq 6$ ) and 14M, an acceleration ratio of more than 10 times was achieved.
Furthermore, the MP approach showed an acceleration ratio of up to 38 times compared to the CPU, which is approximately 3 times faster than the GPU version.
The acceleration ratio is almost saturated on larger systems, implying that the computational resources of GPU are fully utilized.
In contrast, the acceleration ratio of the smaller system is relatively small.
Such low efficiency is due to the inefficient use of GPU computational resources in smaller systems and bottlenecks in parts not yet optimized for GPUs.

It is important to note that the ideal computational performance of four CPU nodes is 20.2 TFLOPS, while that of one GPU node is 38.8 TFLOPS (77.6 TFLOPS with Tensor Core).
Compared to this, our GPU computation speed appears excessive because a speedup of at most 3.8 (77.6/20.2) times is expected, based on a simple ratio from the ideal values.
This is attributed to the fact that we reduce the number of MPI processes to secure sufficient CPU memory. The number of MPI parallelizations on the CPU (128) is 32 times larger than on the GPU (4), leading to bottlenecks due to file input/output (I/O) and MPI communications.
Ignoring the efficiency of thread parallelization, the performance of 128 CPU cores is 5.05 TFLOPS, whereas the performance of the 1 GPU node with Tensor Core is 15.3 times larger than that.

Figure \ref{fig:etime_detail} shows the computation times for the four main tasks.
When comparing the four tasks, the computation time increases in the order of exchange self-energy, core-exchange self-energy, screened Coulomb interaction, and correlation self-energy.
Moreover, the proportion of correlation self-energy calculation increases as the system size increases.
For all of these main tasks, the GPU accelerates calculations, reducing computation time compared to 4 CPU node calculations.
The accelerations by GPU in smaller systems, such as (InAs)$_2$(GaSb)$_2$ and Ni$_2$MnGa NM structure, are relatively lower than those of larger systems.
The reason is that the GPU's computational resources are not fully utilized in smaller systems, and CPU computation and data transfer costs are relatively high.
However, in larger systems, the acceleration ratio of these tasks increases to around 10 and 30 in the GPU and GPU$^\mathrm{MP}$, respectively.
This indicates that GPU resources are utilized more efficiently, and the costs of CPU computation and data transfer are relatively low.

Figure \ref{fig:ctime}(a-d) illustrates the computation time for the kernel parts of each task listed in Table \ref{table:computationalcost}.
$I^\mathrm{c}$ and $R^\mathrm{c}$ correspond to the computational time of the QK$\Omega^\mathrm{I}$ and QK$\Omega^\mathrm{R}$ blocks, respectively, in Algo. \ref{algo:hsfp0}.
As noted in Algo. \ref{algo:hsfp0}, line \ref{line:zwz}, the multiplication of the matrix with $Z$ is shared for $I^\mathrm{c}$ and $R^\mathrm{c}$ parts, thus they are not entirely consistent with Eqs.~(\ref{eq:sigci}) and (\ref{eq:sigcr}).
Note that the data of the calculation of $\epsilon^{-1}$ by GPU$^\mathrm{MP}$ are not listed because we did not employ the MP approach in this part.
As shown in Fig.~\ref{fig:ctime}(a-d), the computation time of these tasks basically increases exponentially, which is described in Table \ref{table:computationalcost}.
Except for $\varSigma^\mathrm{c}$, the calculation of $Z$ takes the most time.
In contrast, for $\varSigma^\mathrm{c}$, the main computation is the imaginary-axis integration ($I^\mathrm{c}$), followed by $Z$ and real-axis integration ($R^\mathrm{c}$).
The GPU and GPU$^\mathrm{MP}$ versions drastically improve the computational time for all kernel parts.
Figure \ref{fig:ctime} (e-l) shows the acceleration ratio between the GPU and the CPU (e-h) and between the GPU$^\mathrm{MP}$ and the GPU (i-l).
Even considering that the current CPU thread parallel efficiency is lower than the ideal value of 4, at approximately 1.5, the GPU acceleration ratio from CPU on $Z$ and $R^\mathrm{c}$ appears excessive.
This is because, as previously mentioned, in the CPU version, the $Z$ computation for $W$ involves MPI communication due to the parallel computation for MPB, and the $R$ computation is limited by the file I/O for $W$.
We assign one MPI process to one GPU, which allows the GPU version to utilize more memory per MPI process than the CPU version, contributing to the excessive acceleration.
In GPU$^\mathrm{MP}$, computations, which mainly consist of matrix multiplication, such as the calculation of $I^\mathrm{c}$, achieve approximately a four-fold speedup compared to GPU.

\subsection{Accuracy of mixed precision calculation}
\begin{figure}[!tb]
  \centering
  \includegraphics[width=8cm]{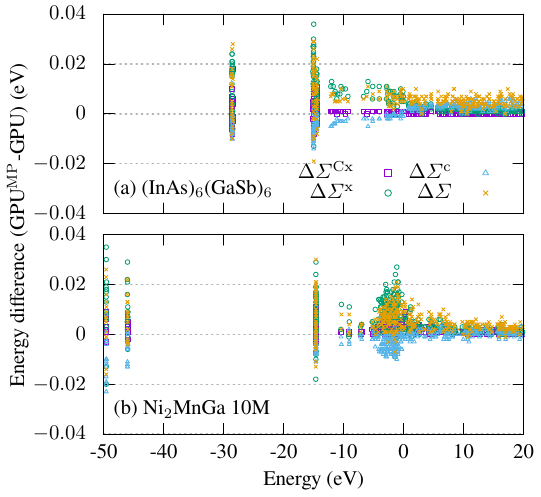}
  \caption{Self-energy and its components differences between MP and DP at $\Gamma$ point, where $\Delta \varSigma^\mathrm{Cx}$, $\Delta \varSigma^\mathrm{x}$,$\Delta \varSigma^\mathrm{c}$ and  $\Delta \varSigma$ are the differences on core-exchange, exchange, correlation, and total self-energies, respectively.
  The horizontal axis represents the quasiparticle energy in LDA/GGA calculations. }
  \label{fig:qpdiff}
\end{figure}
As shown in Fig.~\ref{fig:etime_all}, the computation speed in the MP approach (GPU$^\mathrm{MP}$) is more than three times faster compared to that of the DP approach (GPU).
Although this approach enables faster computations, there are concerns regarding the reduction of the accuracy of the calculations.
To verify the accuracy of the computation results, we compared the quasiparticle self-energy obtained from these approaches.
Specifically, in the first cycle of QSGW calculations, we evaluated the differences in the diagonal components of the self-energies on (InAs)$_6$(GaSb)$_6$ and Ni$_2$MnGa 10M as a representative of each system.

Figure \ref{fig:qpdiff} shows the difference in self energies between MP and DP as a function of the initial eigenvalue.
It shows that the difference between DP and MP varies with energy, tending to be larger at lower energy levels.
Additionally, the difference is smaller in the case of core-exchange self energy.
This may be because the core exchange self-energy calculation uses only the wave functions within the MT spheres, whose basis functions are analytically orthogonal.
The mean differences and their root mean square were \SI{-7}{meV} and \SI{9}{meV} in the (InAs)$_6$(GaSb)$_6$ and \SI{5}{meV} and \SI{6}{meV} in 10M systems, respectively.

\section{Electronic structure from large-scale QSGW calculations}
This subsection addresses the electronic structures obtained from the benchmark calculations.
By considering interfaces and modulation structures, large-scale QSGW calculations enable direct and non-empirical treatment of their contributions.
The DP approach is used for the following results.
The convergence criterion for the QSGW calculations was set to $\Delta_\mathrm{QP}$ = 0.005 eV for quasiparticle energies below 20 eV. 
\subsubsection{Type-II superlattice of (InAs)$_n$(GaSb)$_n$}
\begin{figure*}[tb]
  \centering
  \includegraphics[width=1.0\linewidth]{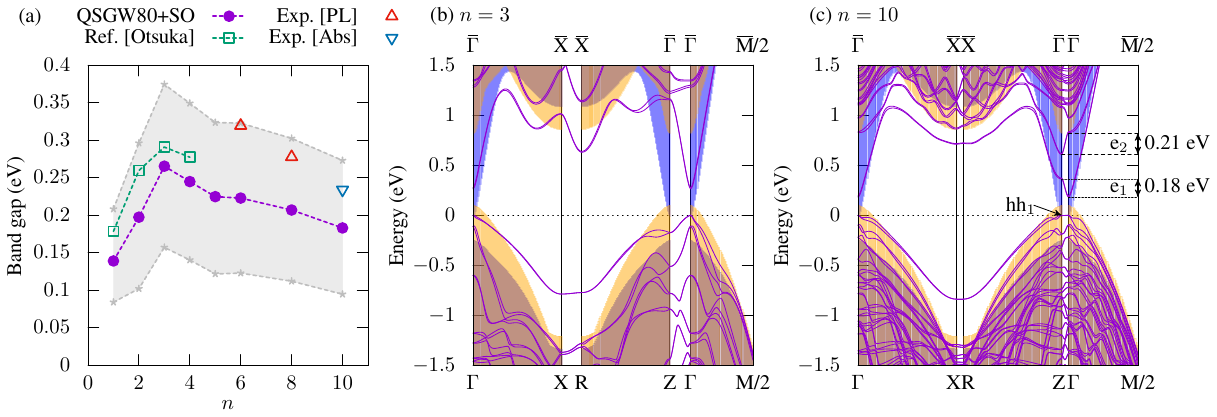}
  \caption{Band gap (a) with respect to $n$. The dashed lines are a guide for the eye.
    Band dispersion of $n=3$ (b) and $n=10$ (c). Ref.~[Otsuka] is \cite{Otsuka2017}, Exp. [PL] and Exp. [Abs] are experimental values from photoluminescence and optical absorption. \cite{Andrew2001, Klein2011}
    These experimental data are extrapolated values to zero kelvin.
  The upper and lower edges of the shadowed area in (a) are the band gap obtained from the non-self-consistent calculation on 70\% and 90\% of hybrid QSGW approaches from the result of QSGW80.
  The shadowed area in (b) and (c) is the projected band dispersion of bulk InAs (blue) and GaSb (yellow) to the two-dimensional BZ (2D-BZ) with band offset correction. The upper horizontal axis indicates the symbols in 2D-BZ.
}
  \label{fig:inasgasb_gap}
\end{figure*}
Figure \ref{fig:inasgasb_gap}(a) shows the band gap obtained by QSGW80 with spin-orbit coupling (SOC) correction.
The results of $n=3$ and $n=5$ are obtained using the same setting of $\vec{k}$ points with $n=4$ and $n=6$, respectively.
We find that the band gap is maximized at $n=3$ and reduced by increasing or decreasing $n$. 
This trend agrees with the previous QSGW80 + SO results of Otsuka et al. \cite{Otsuka2017} and the experimental results \cite{Andrew2001, Klein2011}.
For $n=10$, the band gap value is \SI{0.18}{eV}, a decrease of about 30\% from that of $n=3$ (\SI{0.27}{eV}).
Note that the calculated band gaps for small values of $n$ are slightly lower than those reported by Otsuka \cite{Otsuka2017}, despite both being obtained using the \ecalj package. This discrepancy arises from differences in the numerical treatment of the atomic base set and $\vec{k} \rightarrow 0$ on the $W^\mathrm{c}$, as described in Sec.~\ref{sec:overview_of_qsgw}.

In previous studies using $GW_0$ calculations \cite{Taghipour2018}, the obtained band gap of \SI{0.234}{eV} for $n=10$ is in good agreement with the experimental results of \SI{0.234}{eV} and \SI{0.228}{eV} \cite{Klein2011}. However, it is larger than for $n=8$, contrary to the experimental facts and the present results.
Such discrepancies may be attributed to the fact that the $GW_0$ approach does not update the wave functions from the LDA ones.
The gray area is based on the results of QSGW80 and represents the region enclosed by the band gaps calculated using QSGW70 and QSGW90 with SOC.
The experimental results appear between QSGW80 and QSGW90, implying that the effect of the vertex may not be significant in this material.

The dispersion curves with projected areas of the bulk band are shown in Figs.~\ref{fig:inasgasb_gap}(b) and (c).
We evaluated the band offset to the bulk band of InAs and GaSb based on the energy levels of the core states without SOI, specifically, the Sb $3p$ and As $2p$ states.
Note that bulk bands were calculated with the same in-plane lattice constant as the (InAs)$_n$(GaSb)$_n$ case.
The calculated offset values for the Fermi level of $n=10$ are \SI{0.25}{eV} and \SI{-0.10}{eV} for InAs and GaSb, respectively.
Considering these offsets, the conduction band minimum (CBM) of InAs is higher than the valence band maximum (VBM) of GaSb, indicating the charge transfer from InAs to GaSb by forming the interface.
The obtained value of the difference between the CBM of InAs and the VBM of GaSb is \SI{0.10}{eV}, which is in good agreement with the previous investigation based on experimental measurements of \SI{0.15}{eV} \cite{Vurgaftman2001}.
The quantum well states (e$_1$, e$_2$) are clearly observed in the band dispersion of $n=10$.
The dispersion of e$_1$ and e$_2$ in the Z-$\Gamma$ region is due to the interaction between the quantum well states across the interface.
Since the e$_1$ and e$_2$ states have different parities, their dispersions in Z-$\Gamma$ are reversed.
The bandwidths of e$_1$ and e$_2$ are \SI{0.18}{eV} and \SI{0.20}{eV}, respectively, while the heavy hole state (hh$_1$) exhibits a dispersionless band in the Z-$\Gamma$ region. 
This behavior is also reported with the previous model calculations \cite{Das2022, Mukherjee2021}.
It suggests that, along the transport direction (i.e., $z$-direction), e$_1$ and e$_2$ possess high group velocities, whereas efficient transport of hh$_1$ would require the application of a strong electric field.

\subsubsection{Ni$_2$MnGa martensite phases of NM, 10M, and 14M}
\begin{figure*}[!tb]
  \centering
  \includegraphics[width=0.85\linewidth]{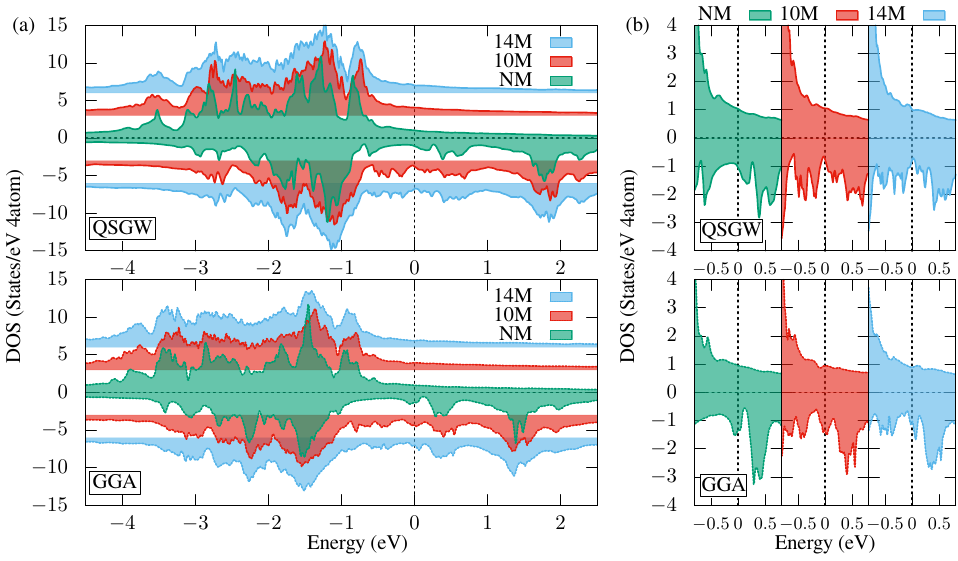}
  \caption{
Total density of states (DOS) (a) and its enlarged plots near the Fermi level (b) of the NM, 10M, and 14M structures on Ni$_2$MnGa.
The origins of the vertical axis of (a) are set as $\pm 3$ and $\pm 6$ for the data of 10M and 14M, respectively.
The top (bottom) graph indicates the results of QSGW (GGA). In all cases, the Fermi energy is set to \SI{0}{eV}. 
}
  \label{fig:ni2mnga_dos}
\end{figure*}
Figure \ref{fig:ni2mnga_dos} shows the density of states (DOS) of Ni$_2$MnGa in the NM, 10M, and 14M phases, assuming a ferromagnetic state.
The results of the PBE version of the generalized gradient approximation (GGA) are also shown for the comparison \cite{Perdew1996}.
By comparison of the QSGW and GGA results, the peak of unoccupied states in minority spins appears around \SI{1.3}{eV} on GGA, whereas it is approximately \SI{0.4}{eV} higher energy in QSGW.
Additionally, the DOS intensity in QSGW is higher, resulting from the narrowing of the bandwidth.
These differences between the QSGW and GGA results arise from the improvement of the excessive delocalization of the $d$ electrons typically observed in GGA.

Focusing on the vicinity of the Fermi level, in GGA, the DOS of minority spin states for all NM, 10M, and 14M structures are located near the peak.
In contrast, in QSGW, the Fermi level is near the valley of the DOS. 
It is at the bottom in the case of 10M and 14M structures, in contrast to the case in NM. 
This implies that the modulated structures observed in 10M and 14M are realized by stabilizing the electronic structure from NM.
The peak structure of the electronic states obtained from the GGA may be suppressed by inducing deformation in the crystal structure.
In fact, it has been reported that GGA overestimates the optimized $c/a$ ratio on the NM structure and energetically stabilizes the artificial structure (4O) compared to 10M and 14M \cite{Zeleny2016}.

One of the reasons for the difference in the electronic structure of minority spins between GGA and QSGW is due to the difference in the magnitude of the exchange splitting.
The calculated magnetic moments per formula unit for the NM, 10M, and 14M structures are 4.60, 4.51, and \SI{4.50}{\bohrmagneton}, respectively, in QSGW. These are in contrast to GGA values, 4.17 (NM), 4.13 (10M), and \SI{4.16}{\bohrmagneton} (14M).
The magnetic moments in QSGW are overestimated compared to the experimental values for the martensite phase: 4.23\cite{Ooiwa1992}, 4.16\cite{Khan2016}, and \SI{4.17}{\bohrmagneton}\cite{Webster1984}.
This increase can be attributed to the enhanced localization of the $d$ electrons, which leads to a larger intra-atomic exchange splitting. Indeed, even within the framework of DFT, the use of advanced exchange-correlation potentials beyond GGA, such as meta-GGA (SCAN) or hybrid functionals (HSE03), has also been reported to yield increased magnetic moments in Ni$_2$MnGa alloy.
Specifically, for the NM structure, magnetic moment of \SI{5.15}{\bohrmagneton} ($c/a$=1.08) with HSE03 and \SI{4.73}{\bohrmagneton}($c/a$=1.17) with SCAN were reported in Ref.~\cite{Janovec2022}, while a value of \SI{4.67}{\bohrmagneton} ($c/a$=1.21)  was obtained with SCAN in Ref.~\cite{Baigutlin2020}.
The overestimation of magnetic moments compared to experimental values is likely due to the absence of spin fluctuations in the calculations \cite{Sponza2017}. Incorporating such effects would be expected to reduce the calculated values.

\section{summary}
We have developed a GPU-accelerated QSGW calculation code by extending the code \ecalj.
This implementation was carried out using OpenACC programming and GPU-accelerated math libraries to maintain the maintainability of computational code.
Two benchmarks have shown that GPU computations successfully accelerated the QSGW calculation, with one GPU node performing calculations equivalent to approximately 40 CPU nodes.
Moreover, in our benchmark, the number of MPI parallelizations in GPU computations was significantly smaller than CPU computations due to the computational performance of the GPU.
This also helped reduce bottlenecks in MPI communication and file I/O, which in turn contributed to the acceleration.
The mixed precision approach was also developed, achieving a speedup of over three times with the double precision version while maintaining acceptable numerical accuracy in self-energy.
The developed GPU-accelerated QSGW code facilitates applications to large-scale systems, such as interfaces, surfaces, and superlattice structures, paving the way to accurate electronic structure calculations in extended systems.

\section{Acknowledgement}
The first-principles quasiparticle self-consistent $GW$ calculations were performed using the Supercomputer Center, Institute for Solid State Physics, University of Tokyo, Japan. 
This work was partially supported by JSPS KAKENHI, Grant-in-Aid for Scientific Research (Nos. 21KK0083, 24K08229, 24K17608, and 25K01665).
The crystal structure illustrations were produced using VESTA \cite{Momma2011}.

\bibliographystyle{elsarticle-num} 
\bibliography{main}

\end{document}